\documentclass[fleqn,usenatbib,useAMS]{mnras}

\DeclareRobustCommand{\VAN}[3]{#2}
\let\VANthebibliography\thebibliography
\def\thebibliography{\DeclareRobustCommand{\VAN}[3]{##3}\VANthebibliography}

\usepackage{graphicx}	
\usepackage{amsmath}	
\usepackage{mathrsfs} 
\usepackage{multicol}        
\usepackage{bm}		
\usepackage{pdflscape}	
\usepackage{placeins}
\usepackage{color}
\usepackage{soul} 
\usepackage{ulem}

\definecolor{lightblue}{rgb}{0.12, 0.56, 1.0}
\definecolor{lightred}{rgb}{1.0, 0.44, 0.44}
\definecolor{Red}{rgb}{0.65,0.08,0.05}
\definecolor{Green}{rgb}{0.15,0.45,0.25}
\definecolor{purple}{rgb}{0.46, 0.3, 0.52}

\usepackage[T1]{fontenc}
\usepackage{ae,aecompl}

\usepackage{newtxtext,newtxmath}

\renewcommand{\thefigure}{\arabic{figure}}

\title[Morphology of dark matter haloes]{Morphology of dark matter haloes beyond triaxiality}

\author[G. Bonnet et al.]{
G. Bonnet,$^{1}$\thanks{E-mail: \href{mailto:guillaume.bonnet@lam.fr}{guillaume.bonnet@lam.fr}}
E. Nezri,$^{1}$
K. Kraljic,$^{1}$
C. Schimd$^{1}$
\\
$^{1}$Aix Marseille Univ, CNRS, CNES, LAM, Marseille, France}

\pubyear{2020}

\begin{document}
\label{firstpage}
\pagerange{\pageref{firstpage}--\pageref{lastpage}}
\maketitle

\begin{abstract}
The morphology of haloes inform about both cosmological and galaxy formation models. We use the Minkowski Functionals (MFs) to characterize the actual morphology of haloes, only partially captured by smooth density profile, going beyond the spherical or ellipsoidal symmetry. We employ semi-analytical haloes with NFW and $\alpha\beta\gamma$-profile and spherical or ellipsoidal shape to obtain a clear interpretation of MFs as function of inner and outer slope, concentration and sphericity parameters. We use the same models to mimic the density profile of $N$-body haloes, showing that their MFs clearly differ as sensitive to internal substructures. This highlights the benefit of MFs at the halo scales as promising statistics to improve the spatial modeling of dark matter, crucial for future lensing, Sunyaev-Zel'dovich, and X-ray mass maps as well as dark matter detection based on high-accuracy data.
\end{abstract}



\section{Introduction}

In the cold dark matter paradigm, initial density perturbations 
are a Gaussian random field with red, almost scale-free power spectrum on scales smaller than $\sim100$~Mpc growing with time, hence driving bottom-up hierarchical clustering.
When the local density attains $\sim5.5$ times the mean matter density of the universe (the exact value depends on cosmology), individual spherical haloes collapse until virialization takes place, then maintaining a universal density profile commonly fitted by NFW \citep{Navarro_1996} or Einasto \citep{Einasto_1965,Wang_2020} models. This process is possibly driven by tidal forces between neighbouring density fluctuations, which induce non-radial motions leading to deviations from sphericity \citep[e.g.][]{Engineer+2000,ShawMota2008}. On more general ground, asphericity already characterizes the initial seeds \citep{EisensteinLoeb1995,ShethMoTormen2001} and might arise during the pre-virialization phase because of tidal interactions, altering the timing of virialization \citep[e.g.][]{White_1984,DelPopoloErcanXia2001}. The overall process produces ellipsoidal or triaxial haloes \citep[e.g.][]{AngrickBartelmann2010}, more frequently prolate with tendency to spherical shape as the mass decreases, as confirmed by numerical simulations \citep[][]{Jing_Suto_2002,Allgood_2006,Hayashi_2007,Vera-Ciro_2011,Zemp_2011,DespaliGiocoliTormen2014,Bonamigo_2015,Vega-Ferrero_2017} and observations of gravitational lensing \citep{Limousin_2013}, Sunyaev-Zel'dovich (SZ) and X-ray galaxy clusters \citep{Sereno_2018}.

At higher level of detail, the morphology of dark matter haloes is more challenging. The process of relaxation is much more complex than described by spherical or ellipsoidal collapse models, especially in the inner halo where the mass assembly history determines the mass and concentration of haloes \citep{LaceyCole1993}. On cluster and galactic scales, the distribution of mass is finally affected by accretion and stripping of small sub-haloes \citep[e.g.][]{Ghigna+1998} and by hydrodynamical processes such as gas cooling \citep[e.g.][]{2004ApJ...611L..73K}, supernova feedback \citep[e.g.][]{Pontzen_2014}, and AGN feedback \citep[e.g.][]{Teyssier_2011}. 
The average value of the so-called concentration parameter, which accounts for the distribution of matter in the core of haloes and mainly depends on the halo mass \citep[e.g.][]{Bullock+2001,2008JCAP...08..006M,2015ApJ...806....4M,Ishiyama_2021}, is controlled by the halo formation epoch \citep{Neto_2007,2012MNRAS.422..185G} and cosmology \citep[][]{Maccio+2008,Kwan+2013}. Its scatter depends on the assembly history, environment, and relaxation state of haloes \citep[e.g.][]{2013MNRAS.434..878S,2014MNRAS.441..378L,2015MNRAS.452.1217C,OmegaWINGS_Biviano+2017}. As shown by \citet{Giocoli+2012}, the mass-concentration relation is finally mainly biased by halo triaxiality, then by the substructures living within the host halo virial radius.

The resulting mass-observable relations underlying X-ray, SZ, and lensing studies of galaxy clusters, which mainly depend on the mass of haloes \citep{2019SSRv..215...25P}, are therefore also largely affected by the internal distribution of matter. Moreover, since a large fraction of massive galaxy clusters are actually identified when they are not fully virialised \citep{Ludlow+2012}, accurate cosmological studies will require an accurate characterisation of their morphology.
Special attention is therefore needed in the analysis of highly spatially-resolved data from powerful instruments such as ROSAT \citep{Kirkpatrick+2021}, XMM-Newton \citep{XXL-Pierre+2017,XXL-Adami+2018,Koulouridis+2021}, eROSITA \citep{Merloni+2020} or Euclid \citep{Laureijs+2011}.

From galaxy clusters to dwarf galaxies, the (sub)halo matter distribution also determines the intensity of dark matter induced $\gamma$-rays or neutrino fluxes. Indeed some dark matter candidate particles might give rise to such high energy emissions through annihilation or decay processes \citep[][]{ Bullock+2001,Sanchez-CondePrada2014,2016ConPh..57..496G}. Its accurate characterisation is needed to maximise the return of observing programs by Fermi-LAT \citep{Acero+2015}, CTA \citep{CTA}, HESS \citep{HESS}, IceCube \citep{IceCube} and KM3NeT \citep{2016JPhG...43h4001A}.

This paper aims at investigating the morphology of dark matter haloes of $\sim10^{13}-10^{15}\mathrm{M}_\odot$ beyond the limit of triaxial symmetry by means of Minkowski functionals (MFs). Introduced in cosmology by \citet{Mecke1994} and extensively used to probe the non-Gaussian morphology of the cosmic microwave background \citep[e.g.][]{SchmalzingGorski1998,Hikage+2008,Planck2015XVII,Buchert_2017} and large-scale structure as traced by galaxy clusters \citep{Kerscher1997}, galaxies \citep{Kerscher_1998,SchmalzingDiaferio2000,Kerscher+2001,Hikage+2003,Kerscher2010,Wiegand_2014,WiegandEisenstein2017}, and neutral hydrogen \citep{Gleser+2006,Yoshiura+2017,Spina2021}, MFs have been more rarely applied to investigate the morphology of isolated structures such as galaxies \citep{Rahman2003,Rahman2004}, galaxy clusters \citep{Beisbart2001,Schimd2021} or superclusters \citep{Einasto_2007_1,Einasto_2008}.
Contrary to parameters such as sphericity, prolateness, elongation, and triaxiality \citep{SpringelWhiteHernquist2004}, which are well-suited for regular (relaxed) clusters, and other statistics used to investigate disturbed (dynamically active) clusters such as halo concentration, centroid shift, power ratio, axial ratio, and position angle \citep{Donahue+2016,Lovisari+2017}, MFs do not rely on any specific symmetry and are mathematically well-grounded, with a clear geometrical and topological interpretation. Moreover, they potentially capture the morphology of substructures hidden by average statistics like the radial density profile.

The rest of the paper is organised as follows. Section~\ref{sec: MFs_review} introduces the theory and interpretation of MFs computed with the germ-grain model (\S\ref{subsec: MF_theory}-\S\ref{subsec: r_halo_filled}), and describes the modelling of haloes by semi-analytical and $N$-body simulations (\S\ref{subsec: halo from profile}-\S\ref{subsec: DEUS}). The analysis of the MFs is discussed in Section~\ref{sec: MFs_impact}, which focuses on structure-less semi-analytical haloes distinguished by spherical or ellipsoidal profile. The Section~\ref{sec: MFs_DEUS_SA} is dedicated to more realistic haloes from $N$-body, showing how much MFs are able to capture the more complex morphology determined by the sub-halo population. Section~\ref{sec: conclu} summarises the results and illustrate the future applications of this study. Computational and numerical aspects of MFs including the issue of sampling and discussions about the region probed by MFs are addressed in four Appendices.

\section{Methods} \label{sec: MFs_review}

\subsection{Minkowski Functionals: germ-grain model for haloes} \label{subsec: MF_theory}

The Minkowski Functionals (MFs) are spatial statistics introduced into cosmology by \citet{Mecke1994} which have been also applied to other fields such as statistical physics  \citep[e.g.][]{Mecke_2000}. 
The MFs are set functionals which generalise the notion of curvature and in dimension $d$ there exist $d+1$ of these functionals.
For $d=3$, the MFs are the volume $V$, the surface area $A$, the integrated mean curvature $H$, and the integrated Gaussian curvature $G$ of continuous bodies. Thanks to the Gauss-Bonnet theorem, the latter is linearly related and usually replaced by the Euler characteristic $\chi$, which is a topological invariant accounting for the number of isolated components, tunnels and cavities according to the formula
\begin{equation} \label{eq: chi}
\chi = \# \mbox{ isolated components } - \#  \mbox{ tunnels } + \# \mbox{ cavities}.
\end{equation}
The usefulness of MFs originates from a Hadwiger theorem \citep{Hadwiger1957}, which ensures that the MFs are the only additive measures that capture all the morphological content of bounded bodies (or regions) while being invariant under translations and rotations of the body, and allowing for progressive approximations of the body. 

More formally \citep{Klain1995,Schneider2013}, the MFs are defined on the convex ring $\mathscr{R} = \{K=\bigcup_{i=1}^N K_i, K_i \mbox{ convex set} \}$, i.e. on every body or bounded region $K$ that can be decomposed into convex sets $K_i\subset\mathbb{R}^d$. The Hadwiger theorem states that any real-valued functional $F: \mathscr{R} \rightarrow \mathbb{R}$ which verifies the following properties:
\begin{itemize}
\item \textit{motion invariance}: $F(gK) = F(K)$ for all set $K \in \mathscr{R}$ and $g \in \mathscr{G}$ (Galilean group);
\item \textit{additivity}: $F(K_1 \cup K_2) = F(K_1) + F(K_2) - F(K_1 \cap K_2)$ for all $K_1, K_2 \in \mathscr{R}$;
\item \textit{continuity}: $\lim_{n\to\infty}F(K_n) = F(K)$ if $\lim_{n\to\infty} K_n = K$ (convergence under Hausdorff measure in $\mathbb{R}^d$);
\end{itemize}
can be written as a linear combination of $d+1$ MFs, $F(K) = \sum_{\mu=0}^d c_\mu V_\mu(K)$, with $V_\mu(K)$ the MFs and $c_\mu$ real-valued coefficients independent of the body $K$. For $d=3$ every real-valued functional satisfying the above properties is then completely characterized by the four MFs.

In cosmological applications, usually concerned by the average morphologies of random fields, MFs per unit volume or MFs densities $v_\mu(K)$ are used instead of the MFs. They allow for comparisons between real processes and analytical models that account for the Poisson, Gaussian, or mildly non-Gaussian processes, as expected at large scales \citep[e.g.][]{1988ApJ...333L..41R,Matsubara2003}. Since our interest lies in the morphology of single haloes, focusing on the scale of galaxy clusters that are highly inhomogeneous spatial objects, we use with raw MFs instead of MF densities, i.e. $V_\mu(K)=\{V,A,H,\chi\}$. Note that several alternative normalisations exist \citep[see][]{Schmalzing1996,Kerscher1997}. As $N$-body simulations deal with haloes as sets of particles, the germ-grain model \citep{Mecke1994} is conveniently used to compute the corresponding MFs. It consists in covering the particles (the germs) with balls of equal radius $r_\mathrm{ball}$ (the grains). The union of these balls forms the continuous body $K$, for which the MFs are well-defined functions of the ball radius $r_\mathrm{ball}$. In this model the MFs capture all information contained in the particle distribution \citep[see e.g.][]{Kerscher_1998}. 

Here the MFs are computed using the code described in \cite{Kerscher1997}. The volume $V$ is computed with a Monte-Carlo algorithm. The three other MFs are computed exactly by summing the local contribution of each individual ball (the so-called partial MFs, see Appendix \ref{subsec: MFs shape}) using the method described in \cite{Mecke1994} \citep[see also][and references therein]{Wiegand_2014}.

\subsection{MFs of haloes }\label{subsec: r_halo_filled}

\begin{figure*} 
 \includegraphics[width=0.9\textwidth]{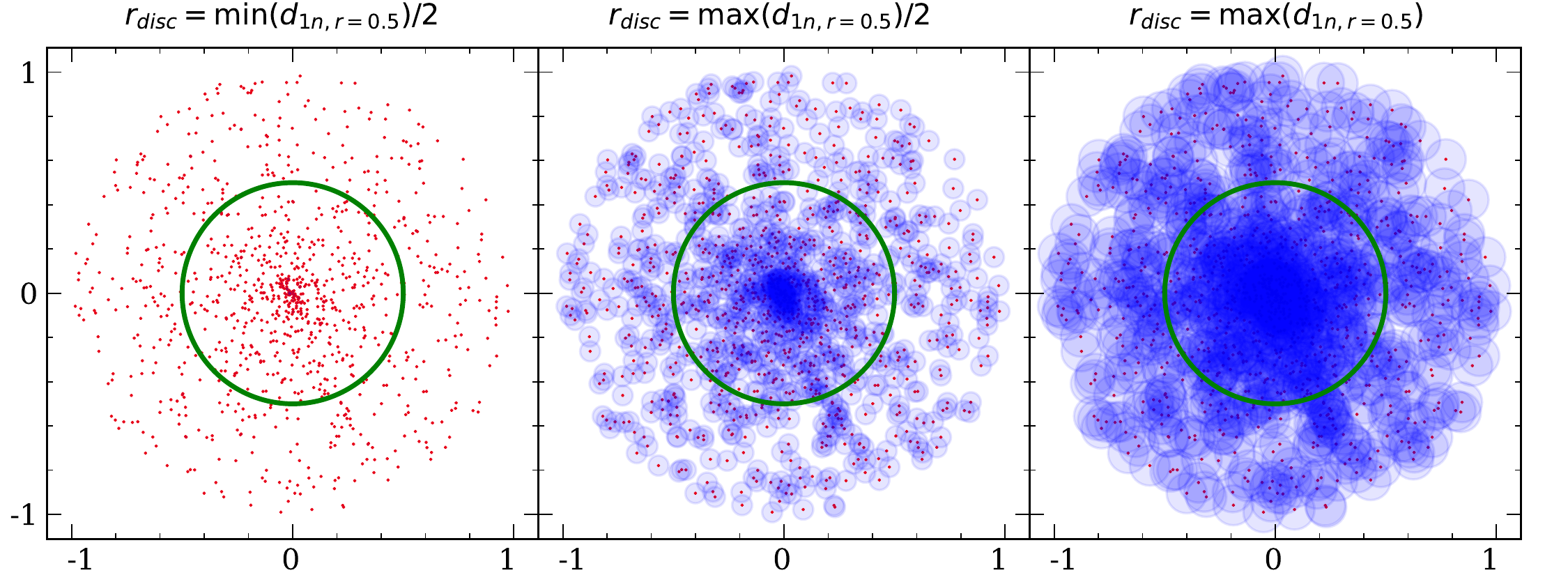}
 \caption{Germ-grain model in two dimensions: particles (red points) are covered by discs (in blue) with three characteristic radii $r_\mathrm{disc}$. Which region of halo is probed by MFs?
 \textit{Left:} the first intersection of two discs occurs for $r_\mathrm{disc} = \min(d_{\mathrm{1n},r=0.5})/2$.
 \textit{Centre:} The more isolated disc touches its first neighbour when $r_\mathrm{disc} = \max(d_{\mathrm{1n},r=0.5})/2$.
 \textit{Right:} For $r_\mathrm{disc} \geq \max(d_{\mathrm{1n},r=0.5})$, the halo inner region is filled and can not change the MFs shape, as most of its particles are covered by their neighbours.
 In 2D/3D, the MFs can probe the morphology of a circle/sphere of radius $r$ in the disc/ball radius range typically given by $[\min(d_{\mathrm{1n},r})/2,\max(d_{\mathrm{1n},r})]$.}
 \label{fig: d_1n_explanation}
\end{figure*}

In the germ-grain model, the MFs exhibit the three following features. First, they depend on the number of particles (see e.g. Equation~\eqref{eq: chi}); as a consequence of additivity, a larger number of particles typically leads to higher amplitudes of MFs (see Appendix \ref{subsec: MF N_part}). Second, the MFs of bounded structures like haloes have similar shape\footnote{Note that the boundary assumed here is the virial radius of \citet{BrianNorman1998}. The halo border is not precisely defined and several definitions exist \citep[see e.g.][]{Knebe_2011,Diemer_2014,Shandarin_2021}. However, the MFs shape remains the same regardless of the definition.} (see Appendix \ref{subsec: MFs shape}); we characterize the location and amplitude of local minima and maxima using idealised haloes generated by semi-analytical models (Section~\ref{sec: MFs_impact}). Third, the radius $r_\mathrm{ball}$ corresponding to extrema decreases with the number of particles, $N$, and is related to the mean interparticle distance $\ell$, which also decreases with $N$. Indeed, the MFs capture the morphological information contained in a region of size $\sim\ell$. For $r_\mathrm{ball}\ll\ell$ balls do not overlap, the MFs are therefore simply proportional to the number of particles; within the virial radius $R_\mathrm{vir}$ containing $N_\mathrm{vir}=N(\leq R_\mathrm{vir})$ particles, one has $V_\mu = N_\mathrm{vir}V_{\mu,1}$ with $V_{\mu,1}=(4\pi r_\mathrm{ball}^3/3,4\pi r_\mathrm{ball}^2,4\pi r_\mathrm{ball},1)$ the MFs of a single ball. For $r_\mathrm{ball}\gg\ell$, only the particles at the halo boundary contribute to the MFs, which depend on the overall ``monolythic'' shape of the body. In particular, for a spherical halo the MFs will be that of a single ball of radius $r_\mathrm{ball} + R_\mathrm{vir}$, while more complex morphologies rarely admit analytical expressions unless considering very specific, ideal shapes \citep{Schimd2021}. See Appendix~\ref{subsec: MFs shape} for more details in terms of local MFs.

In order to assess which part of a halo contribute to the MFs, we consider spherical regions around the centre. When particles of a halo region are entirely covered by neighbouring balls, this region is filled and stops contributing to the total MFs, except for the volume $V$; indeed, $A$, $H$ and $\chi$ take into account only the surface area of uncovered balls (see Appendix \ref{subsec: MFs shape}). Since any realistic halo is typically denser at the centre, germ-grain balls with small $r_\mathrm{ball}$ can fill the inner region while the outer region is still uncovered.
To quantify this dependence, we select a spherical selection of radius $r \leq R_{\mathrm{vir}}$ centred on the halo and containing $N_r = N(< r)$ particles, and compute the distances $d_{\mathrm{1n,} <r}$ of the nearest neighbours for each of these $N_r$ particles.
The halo region corresponding to this selection is considered as filled by balls with radius $r_\mathrm{ball}$ at the radius $r_\mathrm{fill}$ given by
\begin{equation} \label{eq: max_d1n}
r_\mathrm{ball} = \max(d_{\mathrm{1n},<r})\,, \quad\mbox{with}\quad r=r_\mathrm{fill}.
\end{equation}
This equation provides a relation between the MFs and the interparticle distances.

The value of $r_\mathrm{fill}$ computed by Equation~\eqref{eq: max_d1n} delimits the boundary between the contributing and non-contributing regions of the halo to the MFs. This boudary can be actually closer to the centre, i.e. having radius smaller than $r_\mathrm{fill}$, nonetheless this estimation is reasonably close to the exact value as shown in Appendix~\ref{section: halo radius probed}.

A 2D illustration of contributing and non-contributing regions is shown in Figure~\ref{fig: d_1n_explanation}.
Particles (red points) randomly sampling a unit circle with $1/r$ density profile are decorated with discs with increasing radius $r_\mathrm{disc}$ (left to right), the 2D analogue of $r_\mathrm{ball}$.
For any circular region with fixed radius $r$ (thick green circle), if $r_\mathrm{disc}<\min(d_{\mathrm{1n},r})/2$ then the discs are isolated (left panel), while if $r_\mathrm{disc}=\max(d_{\mathrm{1n},r})/2$ then the most isolated particle of the bounded area touches another disc, so that the union of discs in the circle form a connected structure (middle panel). The area of the region bounded by the circle is filled as soon as the discs cover each other; this occurs when Equation~\eqref{eq: max_d1n} is satisfied (right panel).

The filled halo radius $r_\mathrm{fill}$ also informs on a resolution for the MFs. For fixed halo resolution $r_\mathrm{res}$, the MFs are resolved for all the values of $r_\mathrm{ball}$ such that
\begin{equation} \label{eq: res}
r_\mathrm{fill} \geq r_\mathrm{res}
\end{equation}
(remark that $r_\mathrm{fill}$ is function of $r_\mathrm{ball}$). That is, the MFs are not sensitive to morphology of spherical regions of radius smaller than $r_\mathrm{res}$.

\subsection{Semi-analytical haloes}\label{subsec: halo from profile}

\begin{figure*} 
 \includegraphics[width=0.9\textwidth]{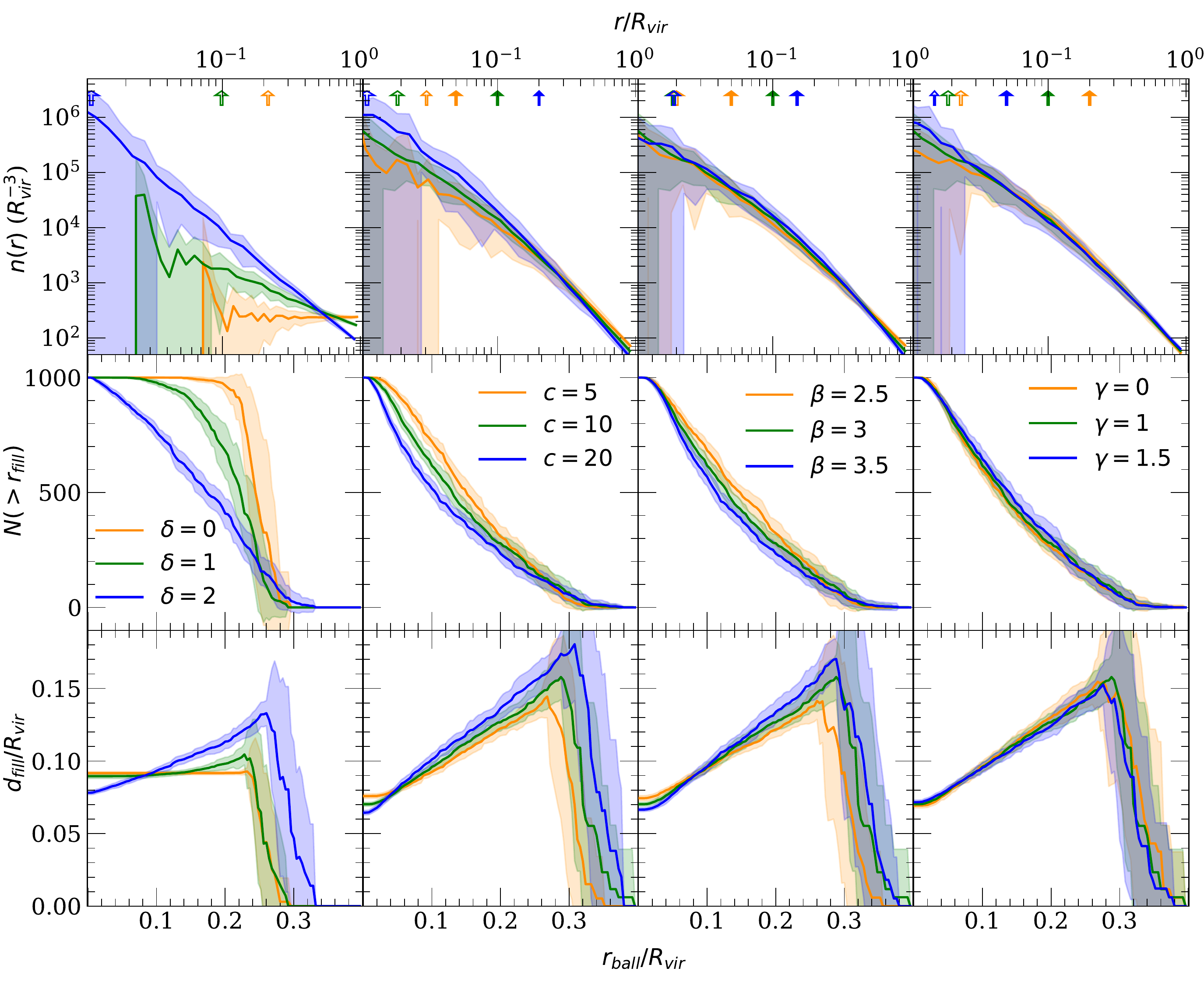}
 \caption{
 \textit{Top panels:} number density $n(r)$ of spherical haloes with scale-free (Equation~\eqref{eq: n(r) d}, first column) and $\alpha\beta\gamma$ profile (Equation~\eqref{eq: n(r) abg}, columns 2-4; all but one parameters allowed to vary, benchmark model with $\beta=3$, $\gamma=1$, i.e. NFW and $c=10$), with $N_\mathrm{sample} = 1000$ particles. The scale radius $r_\mathrm{s}$ and the resolution $r_{10}$ are represented by filled and empty arrows, respectively.
 \textit{Middle panels:} number of particles which contribute to the MFs, $N(>r_\mathrm{fill})$.
 \textit{Bottom panels:} averaged nearest neighbour distances of the contributing particles, $d_\mathrm{fill}$. The MFs amplitude is related to $N(>r_\mathrm{fill})$ because of additivity and the positions of MFs extrema are related to $d_\mathrm{fill}$.} 
 \label{fig: n_r_and_N_r_fill_and_d_fill}
\end{figure*}

To understand the typical shape of MFs of haloes as function of $r_\mathrm{ball}$, we use particle distributions with analytic profile.
This section details the method to generate semi-analytical haloes from exact smooth dark matter distribution and compute the MFs. 
The halo sampling technique is inspired from \citet{Zemp_2011}.
As the MFs are not sensitive to the mass of each individual particle but only to their spatial distribution, we consider the number density profiles $n(r)$. When comparing to a halo with massive particles of equal mass $m_\mathrm{part}$ (e.g. from a $N$-body simulation), the semi-analytical halo is generated with the same total number of particles $N_\mathrm{vir}$.

Two analytical number density profiles are
considered. 
First, a scale-free profile defined by
\begin{equation} \label{eq: n(r) d}
n(r) = \frac{n_0}{r^\delta}
\end{equation}
with constant $\delta$. Note that $\delta=2$ yields a singular isothermal sphere profile, while $\delta=0$ defines a uniform Poisson process, which is routinely used as reference for MF analyses \citep[e.g.][]{Mecke1994}.

The second distribution is the $\alpha\beta\gamma$-profile
\citep{Hernquist_1990,Zhao_1996} defined as
\begin{equation} \label{eq: n(r) abg}
n(r) = n_\mathrm{s} \left( \frac{r}{r_\mathrm{s}} \right)^{-\gamma} \left[ 1 + \left( \frac{r}{r_\mathrm{s}} \right)^{\alpha} \right]^{(\gamma - \beta)/\alpha},
\end{equation}
where $r_\mathrm{s}$ and $n_\mathrm{s}$ are the halo scale radius and density, and $\gamma$ and $\beta$ the (inverse) logarithmic slopes of the inner and outer regions. The parameter $\alpha$ characterizes the transition between inner and outer regions. Note that a NFW density profile \citep{Navarro_1996} corresponds to $(\alpha,\beta,\gamma) = (1,3,1)$ and the scale-free profile is retrieved if $\beta=\gamma$.
Despite the degeneracy between its five parameters \citep{Klypin_2001}, this model allows large flexibility to explore the response of MFs curves. To reduce the dimensionality of the problem, we set $\alpha=1$.
For the $\alpha\beta\gamma$-profile, only the more realistic cases with $\beta>2$ and $0\leq\gamma<2$ are considered. One can further introduce the concentration parameter $c$, which relates the virial radius to the scale radius and the radius where the logarithmic slope is equal to $-2$, i.e.
\begin{equation} \label{eq: r_-2}
c = \frac{R_\mathrm{vir}}{r_{-2}} \hspace{0.5cm} \text{ and } \hspace{0.5cm} r_{-2} = r_\mathrm{s} \left( \frac{2 - \gamma}{\beta - 2} \right)^{1/\alpha}.
\end{equation}

Ellipsoidal haloes are generated by setting the radius in Equation~(\ref{eq: n(r) abg}) to the ellipsoidal form defined as
\begin{equation} \label{eq: r_ellipse}
r_\mathrm{el} = \sqrt{x^2 + \frac{y^2}{Q^2} + \frac{z^2}{S^2}} \quad\mathrm{with}\quad S=\frac{a_3}{a_1}, \quad Q=\frac{a_2}{a_1},
\end{equation}
where $x$, $y$ and $z$ are the Cartesian coordinates along the ellipsoid semi-axis of length $a_1 \geq a_2 \geq a_3$, $S$ is the sphericity and $Q$ the elongation. In order to compare ellipsoidal and spherical haloes, we set $a_1=R_\mathrm{vir}$.

To probe the impact of the parameters $\delta$, $\gamma$, $\beta$, $c$, $S$ and $Q$ on the MFs shape, we generate 30 populations by drawing $N_\mathrm{vir} = 10^5$ particles following the distribution given by Equation~\eqref{eq: n(r) d} or \eqref{eq: n(r) abg} with a fixed choice of parameters), in logarithmically equally spaced shells. Because the computation of MFs for crowded systems like haloes is time-consuming, we randomly subsample these realisations obtaining 30 samples with $N_\mathrm{sample} = 1000$ particles each, allowing us to calculate the mean and standard deviations of MFs.

Figure~\ref{fig: n_r_and_N_r_fill_and_d_fill} shows number density profiles (top panels) of spherical haloes with $N_\mathrm{sample} = 1000$ particles with different values of $\delta$, $c$, $\beta$ and $\gamma$ (see legend). The MFs of the corresponding samples are discussed in Section~\ref{sec: MFs_impact}. 

The NFW density profile gives acceptable fits of the mass distribution of relaxed haloes in cosmological $N$-body simulations at all observable mass range \citep{Wang_2020} and with typical mass-concentration relations \citep{Bullock+2001,Maccio2007,Sanchez-CondePrada2014}. We consider the NFW profile with $c=10$ as benchmark model for a halo of any mass (green curves in the last three columns).

Several conclusions can be drawn :
\begin{itemize}
\item The random selection process might create a sample with a worse final resolution compared to the initial resolution. We therefore use the radius $r_{10}$ of the 10-th innermost particle of the sample as an effective resolution \citep[][see also Appendix~\ref{sec: profile_resolution}]{1985ApJ...298...80C},
\begin{equation} \label{eq: res_SA}
r_\mathrm{res} = r_{10}.
\end{equation}
\item According to Equation~\eqref{eq: max_d1n}, for a fixed radius $r_\mathrm{ball}$ of the germ-grains the MFs are only sensitive to particles outside the sphere with radius $r_\mathrm{fill}$, 
hence the number of particles which contributes to the MFs is
\begin{equation} \label{eq: N_greater_r}
N(>r_\mathrm{fill}) = N_\mathrm{sample} - N(<r_\mathrm{fill}).
\end{equation}
Figure~\ref{fig: n_r_and_N_r_fill_and_d_fill} (central panels) shows $N(>r_\mathrm{fill})$ corresponding to $n(r)$. Since $\delta$, $\beta$ and $c$ (columns 1-3) influence $N(>r_\mathrm{fill})$, they impact the MFs of those samples. The MFs corresponding to larger values of $\delta$, $\beta$ and $c$, which reduce $N(>r_\mathrm{fill})$, are therefore expected to have lower amplitude because of additivity. Note that for all our samples the majority of particles are in the outer region, outside the sphere of radius $r_\mathrm{s}$, and do mostly contribute to the morphology of the halo as described by the MFs.
\item According to the discussion in Section~\ref{subsec: r_halo_filled}, the extrema of the MFs depend on the subset of distances $d_{\mathrm{1n},>r}$ with $r=r_\mathrm{fill}$, which is obtained by removing all the distances corresponding to particles that stop to contribute to the MFs, i.e. at distance $r<r_\mathrm{fill}$.
The averaged nearest neighbour distance of the contributing particles is then defined as 
\begin{equation} \label{eq: d_fill}
d_\mathrm{fill} = \langle d_{\mathrm{1n},>r} \rangle \,, \quad\mbox{with}\quad r=r_\mathrm{fill}.
\end{equation}
Figure~\ref{fig: n_r_and_N_r_fill_and_d_fill} (bottom line) shows $d_\mathrm{fill}$ corresponding to the panels above.
Samples with smaller values of $d_\mathrm{fill}$ correspond to MFs shifted to smaller $r_\mathrm{ball}$. A detailed description of these features is provided in Section~\ref{sec: MFs_impact}.
\end{itemize}

\subsection{N-body haloes: Dark Energy Universe Simulation (DEUS)} \label{subsec: DEUS}

Turning to complex dark-matter haloes we consider the $N$-body haloes from the cosmological simulation DEUS \citep[for details see][]{Rasera_2010,Alimi_2012}. The initial conditions are set with a version of the MPGRAPHIC code \citep{Prunet_2008}. The $N$-body solver is the Adaptive Mesh Refinement RAMSES code \citep{Teyssier_2002}. The dark matter-only simulation is a cubic box of comoving length $L_{\text{box}} = 648 h^{-1}$Mpc, with a comoving spatial resolution $\Delta x = 5 h^{-1}$ kpc. It contains $2048^3$ particles of mass $m_{\rm p} = 2 \times 10^9h^{-1}\mathrm{M}_{\odot}$.
This study is focused on the $\Lambda$CDM simulation, with cosmological parameters matching the analysis of the Wilkinson Microwave Anisotropy Probe Seven-Year data \citep[WMAP7,][]{Spergel_2007} and Union compilation of type Ia supernovae \citep{Kowalski_2008}. This is the reference for the two dynamical dark-energy models examined by the DEUS-FUR suite \citep{Ratra_Peebles_1988}, which will be investigated in a forthcoming paper.

The haloes in DEUS are extraced with a Friend-of-Friend algorithm based on \citet{Roy_2014}. The halo centre is determined as the minimum gravitational potential point $r_\mathrm{cg}$ 
by using the code gyrfalcON \citep{Dehnen2000}. The haloes particles in the full simulation are then selected in a sphere of radius $R_\mathrm{vir}$ according to the \citet{BrianNorman1998} prescription.
The halo resolution is set to $r_\mathrm{res} = 3 \Delta x = 15h^{-1}$kpc.
Similarly to the procedure adopted with semi-analytic haloes, the mean and standard deviation of MFs are computed on 30 random selections of 1000 particles each. Note that we considered only relaxed haloes defined following \citet{Neto_2007}, i.e. having virial ratio $q=2K/|V|<1.35$ and centre-of-mass displacement $s = |\boldsymbol{r}_\mathrm{cg} - \boldsymbol{r}_\mathrm{cm}|/R_\mathrm{vir} < 0.07$, where $K$ and $V$ are respectively the kinetic and potential energy of the halo and $\boldsymbol{r}_\mathrm{cm}$ its centre of mass.

\section{MFs of semi-analytical haloes} \label{sec: MFs_impact}

\subsection{Spherical haloes: impact of $\delta$, $c$, $\beta$, $\gamma$} \label{subsec: SA_spherical}

\begin{figure*} 
 \includegraphics[width=0.9\textwidth]{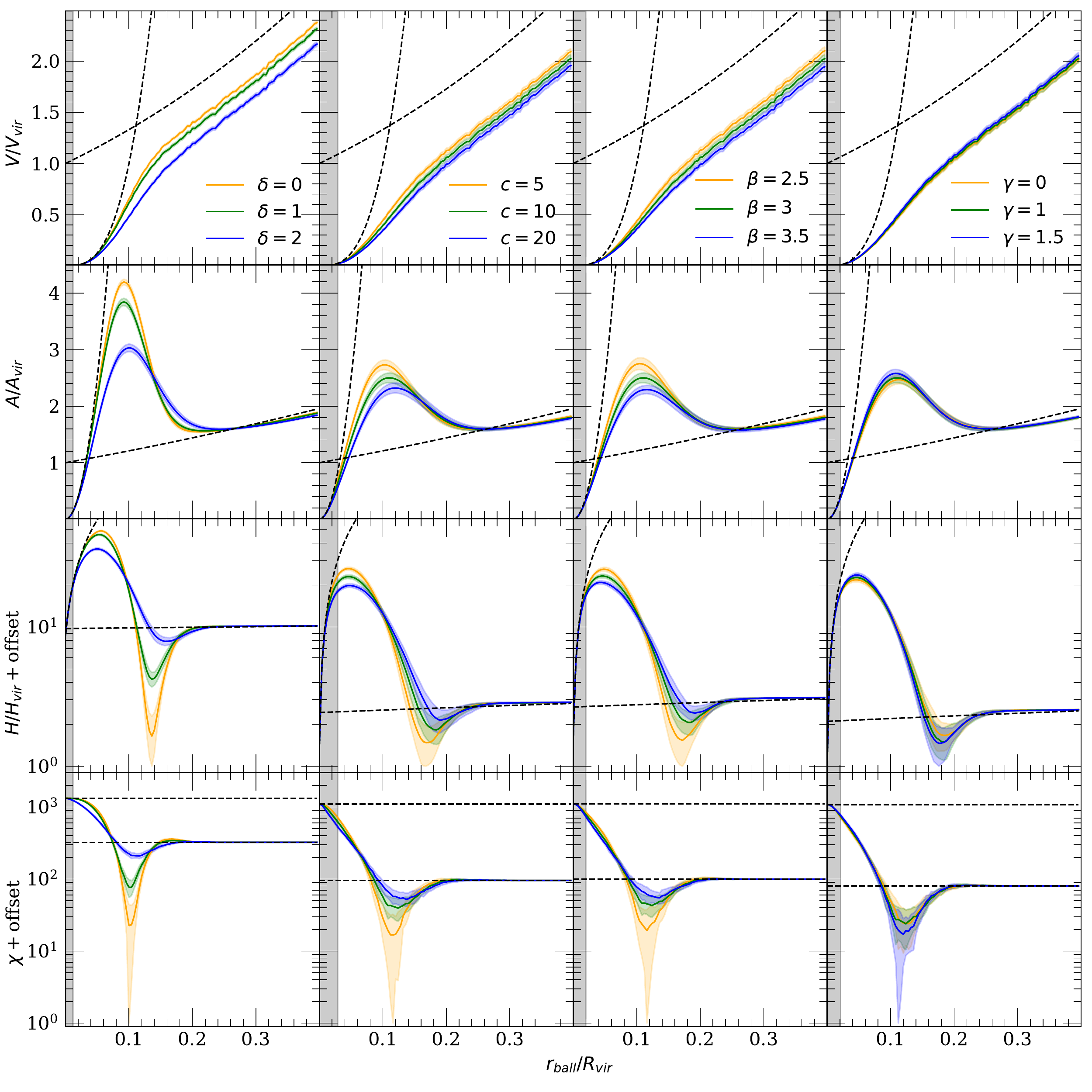}
 \caption{MFs $V_\mu(K)=\{V,A,H,\chi\}$ of the spherical haloes with density profiles shown in Figure~\ref{fig: n_r_and_N_r_fill_and_d_fill}, for different values of $\delta$, $c$, $\beta$, and $\gamma$ (all but one parameter kept fixed).  Error bands represent the standard deviation among 30 realisations of $N_\mathrm{sample} = 1000$ particles each. Values are in units of MFs computed for a ball of radius $R_\mathrm{vir}$; an offset is added allowing the use of logarithmic scale for $H$ and $\chi$. Analytical limits for small and large $r_\mathrm{ball}$ are shown for reference (dashed lines). Grey shaded area defines the resolution limit $r_\mathrm{fill} \leq r_{10}$; see Equation~\eqref{eq: res}. Note that in the first column $r_{10}$ refers to $\delta=2$, as not appropriate for profiles with $\delta=0$ or 1. The amplitude of MFs are determined by their contributing number of particles $N(>r_\mathrm{fill})$ and the $r_\mathrm{ball}$ of the MF extrema are determined by $d_\mathrm{fill}$ (see Figure~\ref{fig: n_r_and_N_r_fill_and_d_fill}).}
 \label{fig: MF_sphere_analytic}
\end{figure*}

Figure~\ref{fig: MF_sphere_analytic} shows the MFs of the semi-analytical haloes considered in Figure \ref{fig: n_r_and_N_r_fill_and_d_fill}, normalised to the MFs for a ball of radius $R_\mathrm{vir}$, i.e. $(V_\mathrm{vir},A_\mathrm{vir},H_\mathrm{vir},\chi_\mathrm{vir})=(4\pi R_\mathrm{vir}^3/3,4\pi R_\mathrm{vir}^2,4\pi R_\mathrm{vir},1)$ (when needed, an offset is added in order to allow the use of a logarithmic scale). 

The concentration parameter $c$ and the slopes $\delta$ and $\beta$ have a similar effect on the MFs. Larger values of these parameters lower the volume $V$, decrease the local maxima of $A$ and $H$, and increase the local minima of $H$ and $\chi$. To smaller extent, the local maximum of $\chi$ also decreases. These trends result from additivity: more concentrated samples, with less particles in the outskirts i.e. smaller $N(>r_\mathrm{fill})$ (see the middle panels in Figure~\ref{fig: n_r_and_N_r_fill_and_d_fill}) yield MFs with smaller amplitude, with lower positive maxima and higher negative minima. The local minimum of the area is an exception; it slightly increases with $\delta$ and $c$, while larger values of these parameters imply fewer contributing particles (see Appendix~\ref{subsec: MFs shape}). 

The parameters $\delta$, $c$, and $\beta$ also influence the position of the MFs extrema $r_\mathrm{ball}$, albeit in more subtle way. As discussed in Section~\ref{subsec: r_halo_filled} and shown in Figure~\ref{fig: n_r_and_N_r_fill_and_d_fill}, for higher values of $\delta$, $c$ and $\beta$ the distance $d_\mathrm{fill}$ is smaller (larger) at small (large) $r_\mathrm{ball}$. The MFs are shifted accordingly. For example, the position of the maximum of $H$ increases with concentration $c$ (panel in bottom line and second column of Figure~\ref{fig: MF_sphere_analytic}) consistently with the maximum of $d_\mathrm{fill}$ for the same value of $r_\mathrm{ball}$ (Figure~\ref{fig: n_r_and_N_r_fill_and_d_fill}, second column).

The response of MFs to the inner slope $\gamma$ follows a weaker and opposite trend to the change relative to $\delta$, $c$ and $\beta$, as expected from $N(>r_\mathrm{fill})$; see panels of second line in Figure~\ref{fig: n_r_and_N_r_fill_and_d_fill}. When $r_\mathrm{fill} \gtrsim r_\mathrm{s}$, the inner region becomes filled and the inner slope no longer impacts the MFs. Since the MFs reach their extrema for $r_{\rm ball}$ such that $r_\mathrm{fill} \gtrsim r_\mathrm{s}$, the different behaviours observed for these points do not depend on the inner slope. However, higher values of $\gamma$ correspond to lower values of $r_\mathrm{s}$ (with $\alpha$, $\beta$ and $r_{-2}$ kept constant, see Equation \eqref{eq: r_-2}) and thus to lower outer slope at a given halo radius $r$. The indirect impact of $\gamma$ is therefore the opposite of $\beta$. Although the MFs do not depend on the halo inner region, by varying the radial extension of the sample selection it is possible to probe this inner region and thus the direct impact of $\gamma$ (see Appendix~\ref{sec: MFs probe different region}).

\subsection{Ellipsoidal NFW haloes: impact of $S$ and $Q$} \label{subsec: MF ell}

\begin{figure*} 
 \includegraphics[width=0.9\textwidth]{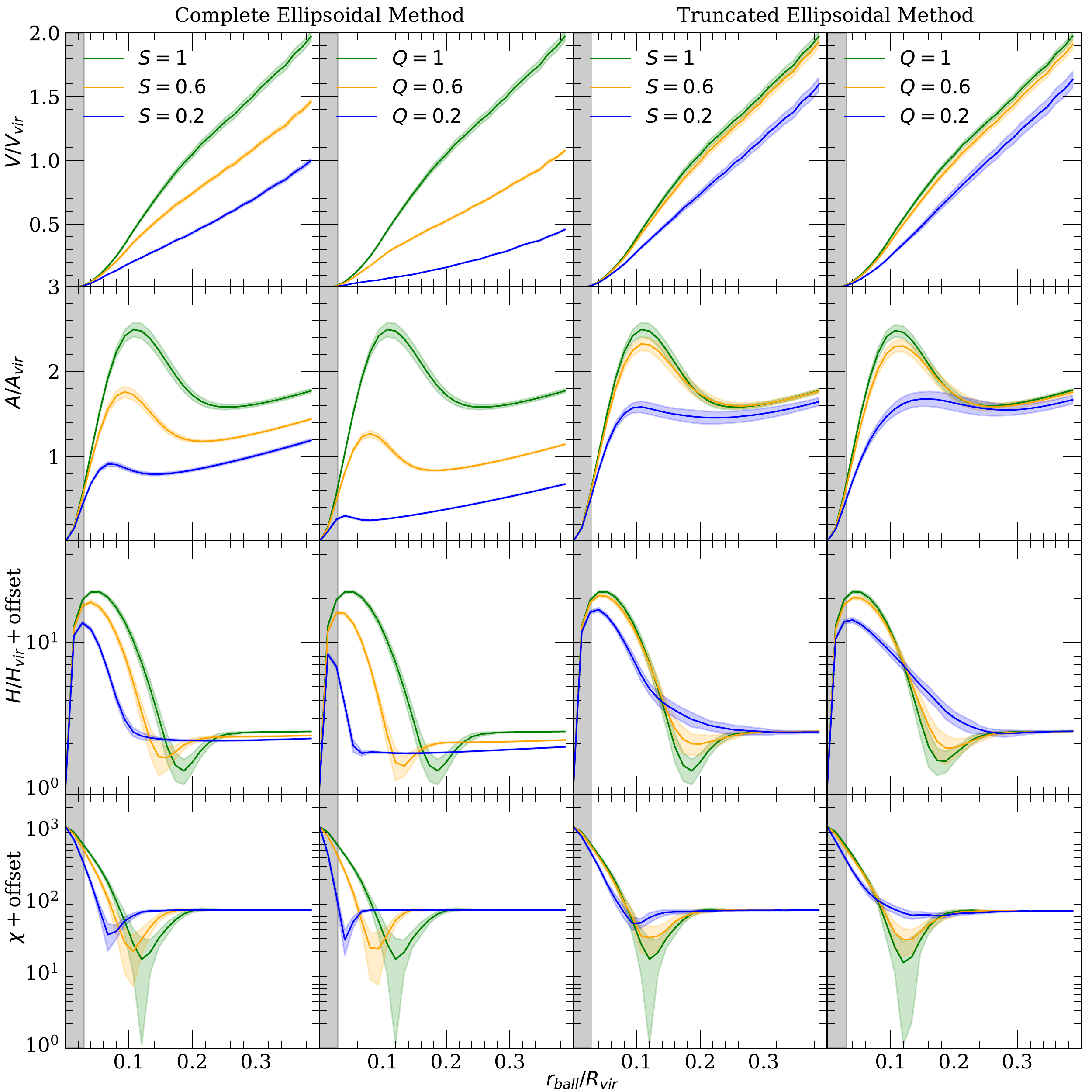}
 \caption{MFs of NFW ellipsoidal haloes. Same conventions as in Figure~\ref{fig: MF_sphere_analytic}. Columns 1 and 2 (3 and 4) refer to oblate and prolate haloes, generated using the Complete (Truncated) Ellipsoidal Method. Less spherical haloes have smaller size thus their MFs amplitudes are both smaller and shifted to smaller $r_\mathrm{ball}$.}
 \label{fig: MF_ellipsoidal_analytic}
\end{figure*}

The two methods are considered for generating ellipsoidal haloes: 
\begin{itemize}
\item Complete Ellipsoid Method: particles are generated in ellipsoidal shells until $r_\mathrm{el} \leq R_\mathrm{vir}$. This way, the sphere of radius $R_\mathrm{vir}$ does contain empty spaces and all the ellipsoidal shells are complete. Note that no spherical selection is performed here. This method mimics the result of an iso-density (or iso-potential) selection applied on a halo with an ellipsoidal shape obtained from a $N$-body simulation, thus conserving the natural shape of the halo.
\item Truncated Ellipsoid Method: particles are generated in ellipsoidal shells until $r_\mathrm{el} \leq R_\mathrm{vir}/S$. Then the particles are selected inside a sphere of radius $R_\mathrm{vir}$. This way, the spherical selection does not contain empty spaces and the outermost shells are truncated. This method mimics the results of a spherical selection applied on a halo with ellipsoidal shape generated by a $N$-body simulation (the most frequently used in the literature).
\end{itemize}

For both methods and many pairs ($S,Q$), the MFs are computed as explained in Section~\ref{subsec: halo from profile}. Figure~\ref{fig: MF_ellipsoidal_analytic} shows resulting curves for NFW ellipsoidal haloes with $c=10$. The MFs of oblate ($Q=1 \geq S$) and prolate ($1 \geq Q = S$) haloes have larger amplitude and are shifted towards larger $r_\mathrm{ball}$ for larger $S$ and $Q$, regardless the use of the Complete Ellipsoidal Method (columns 1-2) or the Truncated Ellipsoidal Method (columns 3-4). This is a consequence of the halo size, ellipsoidal haloes being smaller than spherical ones. The Complete Ellipsoidal Method produces haloes  where this effect is stronger, as they are not the result of a spherical selection, especially for prolate haloes which are smaller than oblate haloes with same sphericity $S$. This only holds at small $r_\mathrm{ball}$ for the Truncated Ellipsoidal Method, as long as the MFs are dominated by particles located inside the complete ellipsoid i.e. with $r_\mathrm{el} \leq R_\mathrm{vir}$. For germ-grain balls with larger radius, the halo inner region is filled and the shape formed by the contributing particles is not anymore an ellipsoid of parameters $S$ and $Q$, but rather a truncated sphere that better approximates a sphere for prolate than oblate haloes. Note that the interpretations in term of $N(>r_\mathrm{fill})$ and $d_\mathrm{fill}$ considered in Section~\ref{subsec: SA_spherical} applies also here. For the sake of comparison, MFs of ellipsoidal haloes generated with NFW profiles for different concentrations are shown in Appendix~\ref{sec: MFs_SQ_c}. 

\section{Comparison with N-body haloes}\label{sec: MFs_DEUS_SA}

\subsection{Deviation from smooth triaxiality: a case study}

\begin{figure} 
 \includegraphics[width=\columnwidth]{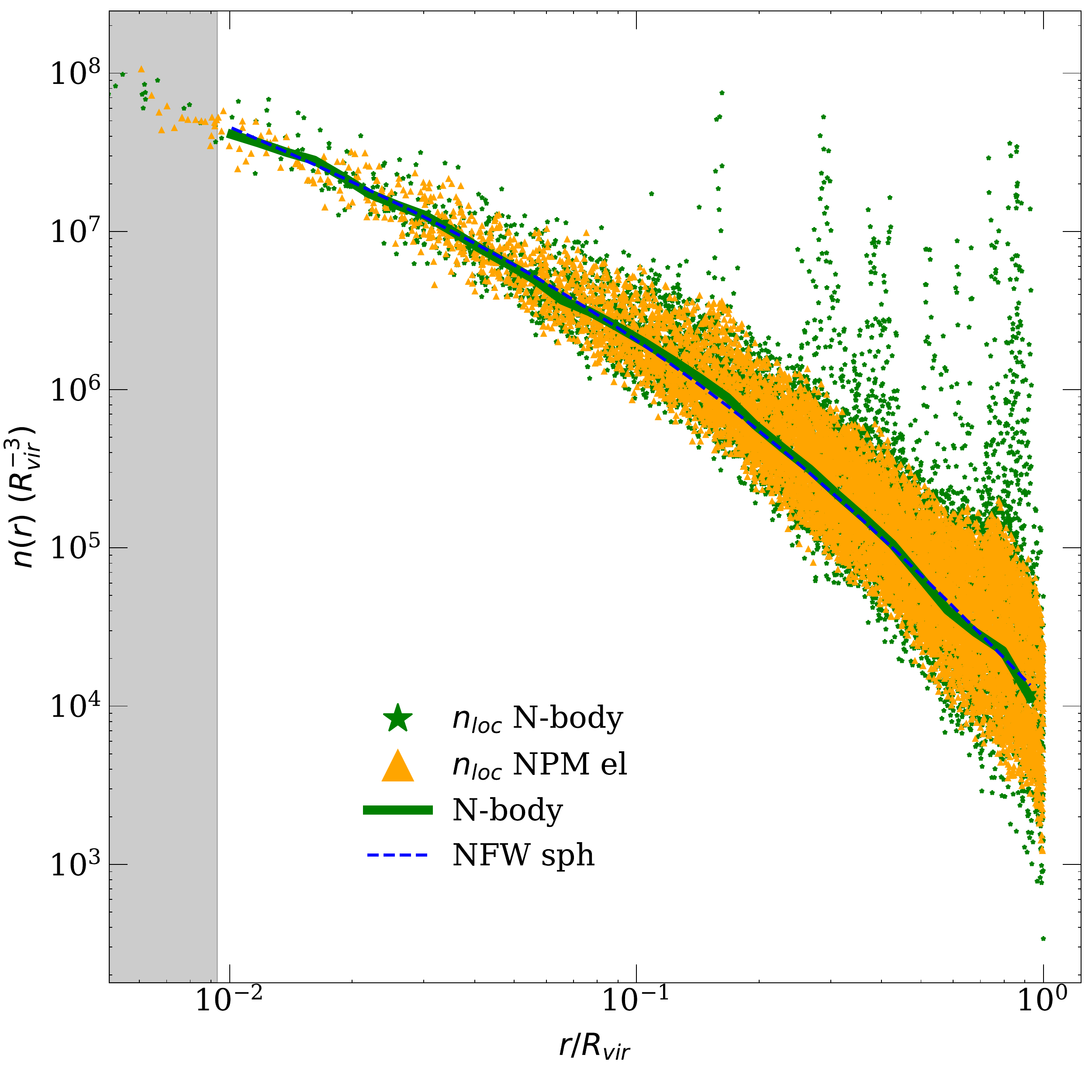}
 \caption{Density profile $n(r)$ and local density $n_\mathrm{loc}(\boldsymbol{r})$ of a $N$-body halo (thick green solid line and symbols). For comparison, $n(r)$ for a semi-analytic halo generated with the parametric method using a spherical NFW profile (``NFW sph''; thin blue dashed line) and $n_\mathrm{loc}(\boldsymbol{r})$ of an ellipsoidal halo generated with the non-parametric method (``NPM el''; orange symbols). The grey shaded area defines the resolution limit, $r_\mathrm{rec}=15 h^{-1}$ kpc.}paper
\label{fig: n_r_n_loc}
\end{figure}

\begin{figure*}
 \includegraphics[width=0.8\textwidth]{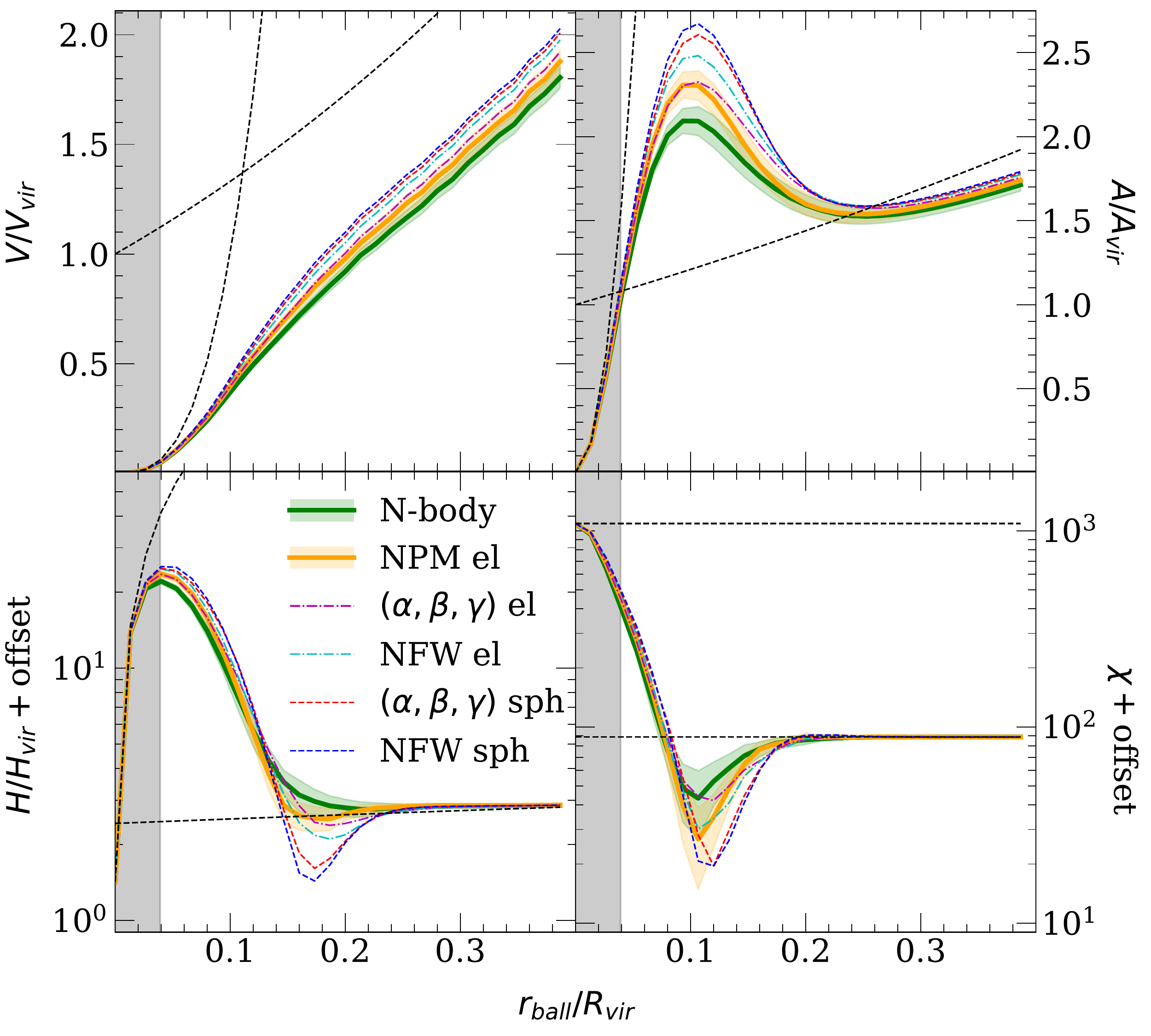}
 \caption{MFs of haloes considered in Figure~\ref{fig: n_r_n_loc} (same colour code) and MFs of haloes generated with the parametric method (other thin lines). The standard deviations are not shown for the MFs of haloes generated with the parametric method (all the thin lines) for a better visibility, but they are similar for all the curves. Same conventions as in Figure \ref{fig: MF_sphere_analytic}. The grey shaded area corresponds to the maximum of all the uncertain $r_\mathrm{ball}$ obtained with Equations~(\ref{eq: res},\ref{eq: res_SA}) and $r_\mathrm{rec}=15h^{-1}$kpc. Haloes generated with $\alpha\beta\gamma$-profile and/or with ellipsoidal shape have MFs closer to the $N$-body halo than those generated with NFW and/or spherical shape, but even for the non-parametric smooth ellipsoidal halo (orange) the MFs are significantly different from the $N$-body halo (green). Thus the MFs are sensitive to more complex morphological features than contained in smooth profiles and are a promising probe for substructures.}
 \label{fig: MF real vs SA}
\end{figure*}

This section demonstrates how much MFs are able to characterize the dark matter distribution in haloes beyond the spherical or ellipsoidal averaged density and thus account for anisotropies beyond triaxial shape and substructures. Substructures account for both subhaloes and tidal streams, i.e disrupted subhaloes with still distinctive signatures in phase-space compared to particles in the main halo \citep[][]{Johnston_1996,Helmi_2003}.

In order to compare the MFs of $N$-body and semi-analytical haloes, we proceed as follows. First, the number density profiles of $N$-body haloes are estimated by counting the number of particles in shells. Then, for each $N$-body halo we generate spherical or ellipsoidal semi-analytical haloes from smooth realistic density profiles following two methods:
\begin{itemize}
\item parametric method:
particles are drawn in shells following the distribution given by Equation~\eqref{eq: n(r) abg}, with  parameters $c$, $\beta$ and $\gamma$ fixed by fitting the density profile of the $N$-body halo;
\item non-parametric method (NPM):
particles are drawn in shells following the distribution induced by the number density profile of the $N$-body halo.
\end{itemize}
Note that for ellipsoidal haloes one first computes the parameters $S$ and $Q$ of the $N$-body halo and then replace the spherical radius by the ellipsoidal radius $r_{\rm el}$ defined in Equation~\eqref{eq: r_ellipse}. Moreover, if haloes are generated with the non-parametric method, we ensure that both their spherical and ellipsoidal density profiles $n(r)$ and $n(r_{\rm el})$ are identical to those of the $N$-body haloes. The non-parametric ellipsoidal density profiles are the most realistic ones. The MFs are finally computed on subsamples of particles randomly selected from the parent haloes, as discussed in Section \ref{subsec: halo from profile}.

For illustration, Figure~\ref{fig: n_r_n_loc} shows the spherical number density profile $n(r)$ of a relaxed halo from DEUS (solid green line) with mass $M_\mathrm{vir} \sim 10^{14.6}h^{-1}M_{\odot}$, similar to the Coma cluster \citep{Gavazzi_2009}, and selected as detailed in Section~\ref{subsec: DEUS}. The density profile of a semi-analytical halo generated with a NFW profile (blue dashed thin line) only slightly differs from that of the $N$-body halo, as expected for a relaxed halo \citep{Neto_2007}. These differences are even smaller for a halo generated with $\alpha\beta\gamma$-profiles and disappear for haloes generated with the non-parametric method (not shown). However, the local density $n_\mathrm{loc}$ of semi-analytical haloes and the $N$-body halo, estimated around each individual particle using the gyrfalcON code \citep{Dehnen2000}, still differ. The local density of the $N$-body halo (green stars) shows localised excess or spikes caused by substructures, especially at large radii ($r>0.1 R_\mathrm{vir}$). Spikes are instead absent  in the ellipsoidal halo generated with the non-parametric method (orange triangles) and in spherical haloes generated with parametric method and NFW or $\alpha\beta\gamma$-profiles, which are smooth by construction. Therefore, smooth density profiles do not probe all the morphological features of $N$-body haloes.

To quantify how much MFs are able to capture and highlight such information, we estimate the statistical difference between the average MFs of $N$-body haloes ($\bar{V}_{\mu,\mathrm{N-body}}$) and semi-analytical haloes ($\bar{V}_{\mu,\mathrm{SA}}$) generated with smooth density profiles by
\begin{equation} \label{eq: chi_2}
\chi_\mu^2 = \frac{1}{N_\mathrm{b}} \sum_{r_\mathrm{ball}} \frac{\Delta V_\mu^2}{  \sigma_\mu^2},
\end{equation}
with 
\begin{align}\label{eq: mean_diff-sigma}
\Delta V_\mu & =  \bar{V}_{\mu,\mathrm{SA}}(r_\mathrm{ball}) - \bar{V}_{\mu,\mathrm{N-body}}(r_\mathrm{ball}) \\ 
\sigma_\mu^2 & =  \sigma_{\mu,\mathrm{SA}}^2(r_\mathrm{ball}) + \sigma_{\mu,\mathrm{N-body}}^2(r_\mathrm{ball})
\end{align}
and where the sum is calculated over $N_\mathrm{b}$ values of $r_\mathrm{ball}$. For each halo, the mean and the standard deviation of the MFs are computed on 30 samples as described in Section~\ref{subsec: halo from profile}.
Figure~\ref{fig: MF real vs SA} shows the MFs of both the $N$-body halo and the two semi-analytical haloes considered in Figure~\ref{fig: n_r_n_loc} (same colour code), as well as the MFs of spherical and ellipsoidal semi-analytic haloes generated with the parametric method based on NFW and $\alpha\beta\gamma$-profiles (other thin lines). The MFs shapes of all these haloes are qualitatively very similar. They exhibit extrema in $A$, $H$ and $\chi$, with amplitude and position depending on the number of balls as shown in Appendix~\ref{subsec: MF N_part}. These extrema result from the increase of both the ball size and the fraction of particles in the filled region (see Figure~\ref{fig: d_1n_explanation}), as this filled region does not contribute to $A$, $H$ and $\chi$ (see details in Appendix~\ref{subsec: MFs shape}). Note that for $H$ and $\chi$, they also result from additional contributions due to the intersection of balls \citep[see Equation~(29) in][]{Mecke1994}.
As expected, haloes generated with more realistic density profiles, i.e. $\alpha \beta \gamma$-profiles (magenta/dot-dashed and red/dashed lines) instead of NFW (blue/dot-dashed and cyan/dashed), have MFs closer to the MFs of the $N$-body halo. Moreover, for fixed parameterization of the density profile, the MFs of haloes generated with the more realistic ellipsoidal shapes (dot-dashed lines) are closer to the MFs of the $N$-body halo than the MFs of haloes with spherical shape (dashed lines). Besides, note that the MFs of a spherical halo generated with an $\alpha\beta\gamma$-profile (red line) differ from the MFs of the $N$-body halo more than the MFs of an ellipsoidal halo generated with a NFW profile (cyan line). This aspect is expected for relaxed haloes in the mass range we consider, which are typically very triaxial \citep[see e.g.][]{Allgood_2006}. Indeed, their profile parameters differ mildly when fitted with NFW or $\alpha\beta\gamma$-profiles. 

To summarize, the MFs of all the semi-analytical haloes significantly differ from the MFs of the $N$-body halo ($\chi_\mu^2 > 1$ for at least one of the four MFs), proving that the MFs are definitely sensitive to morphological information beyond what is captured by smooth triaxial density profiles.

\subsection{Mass and relaxation state: statistical analysis}

\begin{figure} 
 \includegraphics[width=\columnwidth]{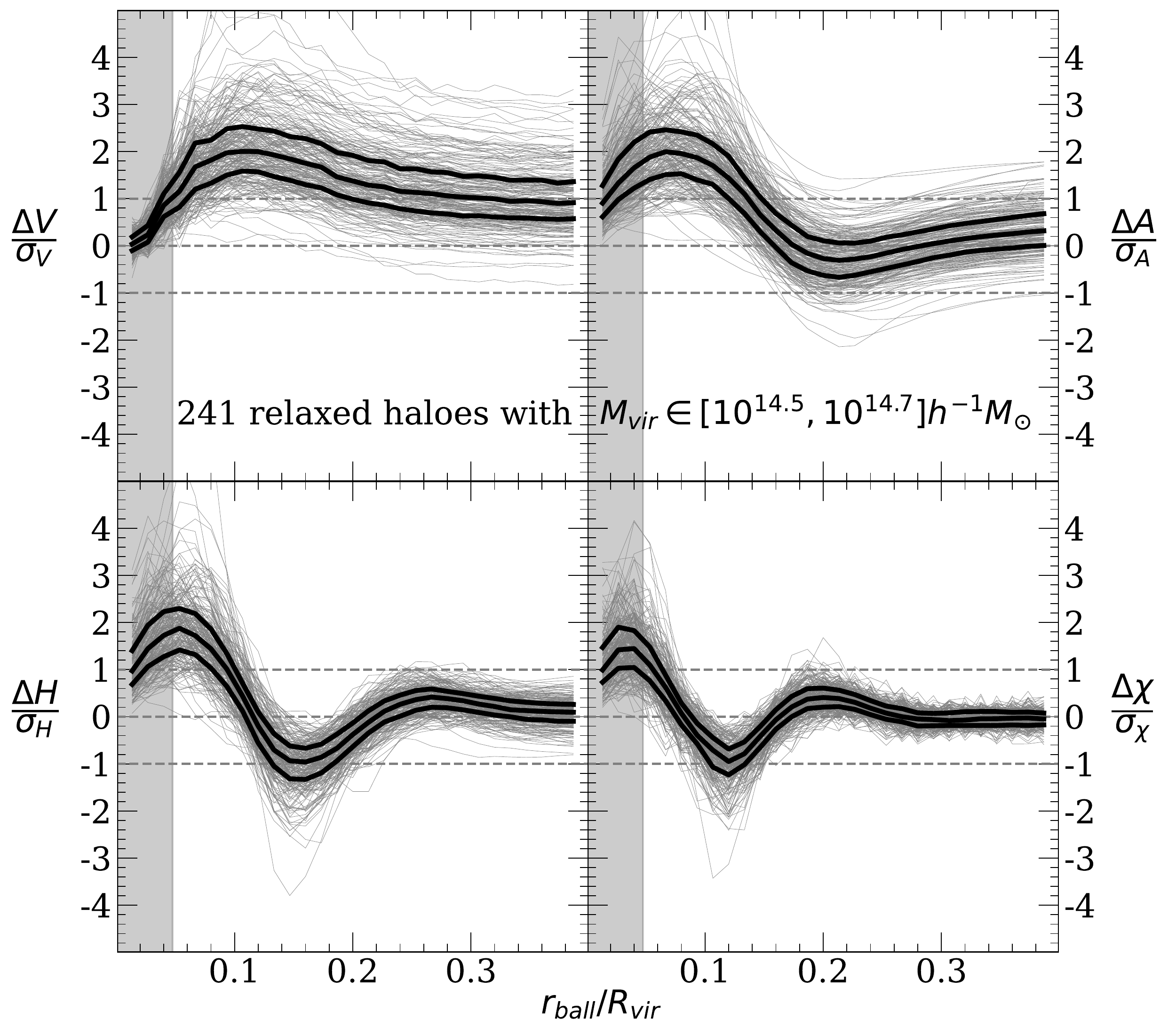}
 \caption{Deviation from smooth triaxiality detected by MFs. Statistical difference $\Delta V_\mu/\sigma_{V_\mu}$ (Equation~\eqref{eq: mean_diff-sigma}) between $N$-body haloes and the corresponding ellipsoidal haloes generated with the non-parametric method. Thin grey lines correspond to 241 relaxed haloes with mass $10^{14.5}<M_\mathrm{vir}h/\mathrm{M}_\odot<10^{14.7}$; black lines mark median and quartiles. The grey shaded area corresponds to the value of resolution as in Figure~\ref{fig: MF real vs SA}. The substructures are expected to cause the observed differences, which are significant ($\chi_\mu^2 > 1$) for more than 80 percent of the haloes.
 }
\label{fig: MF_SA_vs_DEUS_many}
\end{figure}

The comparison between ellipsoidal semi-analytic haloes generated with the non-parametric method and $N$-body haloes is eventually repeated on all the 241 relaxed haloes of the DEUS simulation considered, in the mass range $10^{14.5} < M_\mathrm{vir} h/\mathrm{M}_\odot < 10^{14.7}$. More than 80 percent of these haloes gives $\chi_\mu^2 > 1$ for at least one of the four MFs.\footnote{The substructure mass fraction criterion for the relaxed haloes selection is not taken into account in this study, our goal being to highlight the sensitivity of MFs to substructures. According to \citet{Neto_2007}, the fraction of relaxed haloes should be higher than 50 percent when this criterion is taken into account, which leads to at least 78 percent of relaxed haloes with $\chi_\mu^2>1$ in this mass range.} The differences are typically higher for the MFs with higher order (lower $\mu$), on average $\chi_0^2 \sim 2 \chi_1^2 \sim 3 \chi_2^2 \sim 7 \chi_3^2$. In order to evaluate the ball radius at which significant statistical differences occur, we compute $\Delta V_\mu/\sigma_\mu(r_\mathrm{ball})$ for these haloes as shown in Figure~\ref{fig: MF_SA_vs_DEUS_many}.
The differences are typically positive around the MFs maxima and negative around the MFs minima meaning that the MFs of semi-analytical haloes have higher amplitude than those of $N$-body ones, a trend that is very likely due to substructures. A halo with substructures has indeed typically less contributing particles in the outskirts, namely lower $N(>r_\mathrm{fill})$, than a smooth halo with the same density profile. Some of its particles are inside these substructures and are thus recovered by the balls of their neighbours. As shown in Section~\ref{subsec: SA_spherical}, haloes with lower $N(>r_\mathrm{fill})$ typically have MFs with lower amplitudes, the existence of substructures therefore results in MFs with lower amplitudes, as observed in Figure~\ref{fig: MF_SA_vs_DEUS_many}. We have checked that similar trends appear when including haloes with subhaloes in a simple semi-analytical test (see Appendix~\ref{sec: MFs_sub}).

We checked the ability of MFs to probe the morphology of relaxed and unrelaxed halos with different masses, $10^{12.5} < M_\mathrm{vir} h/\mathrm{M}_\odot < 10^{12.7}$, $10^{13.5} < M_\mathrm{vir} h/\mathrm{M}_\odot < 10^{13.7}$ and $10^{14.5} < M_\mathrm{vir} h/\mathrm{M}_\odot < 10^{14.7}$. Significant differences ($\chi_\mu^2 > 1$) are found for about 95 percent of unrelaxed haloes  in all three considered mass bins. As expected, the differences between the MFs of unrelaxed $N$-body haloes and the MFs of semi-analytical haloes are even more important than for the relaxed haloes. Indeed, the unrelaxed $N$-body haloes have typically experienced more recent major merger and are thus not well-modelled by smooth profiles. Besides, the fraction of relaxed haloes with significant differences between their MFs and the MFs of semi-analytical haloes decreases to 50 percent in the two lower mass ranges. This is expected as related to numerical limitations regarding resolution and disruption of substructures in $N$-body simulations \citep[e.g.][]{van_den_Bosch_2018_sim,van_den_Bosch_2018_SA}.

\subsection{Impact of systematics}

\begin{figure} 
 \includegraphics[width=\columnwidth]{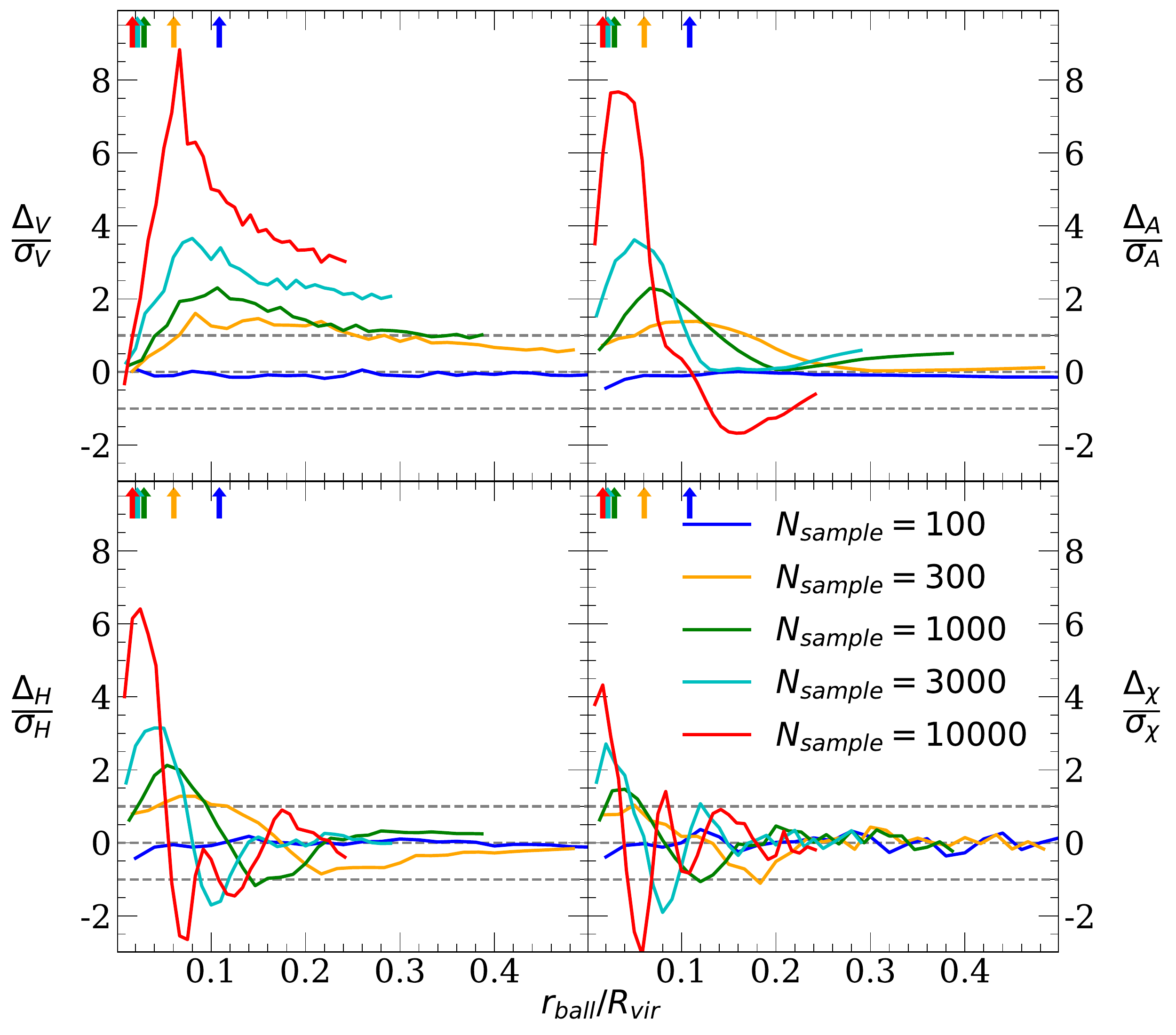}
 \caption{Effect of sampling on MFs differences. Statistical difference $\Delta V_\mu/\sigma_{V_\mu}(r_\mathrm{ball})$ between the MFs $V_\mu$ of the $N$-body halo of Figure~\ref{fig: MF real vs SA} and the MFs of the semi-analytical halo generated with the non-parametric method for different values of $N_\mathrm{sample}$. Arrows indicate the resolution computed with Equation~\eqref{eq: res}.}  
\label{fig: MF_N_body_SA_N_part}
\end{figure}

It is worth to evaluate how much the MFs are sensitive to the sample particle number. Figure~\ref{fig: MF_N_body_SA_N_part} shows $\Delta V_\mu/\sigma_\mu(r_{\rm ball})$ of the $N$-body halo with MFs shown in Figure~\ref{fig: MF real vs SA} and there computed for a sub-sample with $N_\mathrm{sample}=1000 $ particles, now resampled with $N_\mathrm{sample}=100, 300, 1000, 3000, 10000$. The differences between the MFs of the $N$-body and the semi-analytical halo are significant for $N_\mathrm{sample} \gtrsim 1000$.
As expected, these differences increase with increasing $N_\mathrm{sample}$ as the underlying particle distribution in the $N$-body halo is better retrieved while the intrinsic scatter of MFs due to the sampling decreases. The impact of $N_\mathrm{sample}$ on the MFs $V_\mu$ is discussed in Appendix~\ref{subsec: MF N_part}.

Finally, we consider the impact of the sphericity and the elongation of the more massive $N$-body haloes when computed as function of the halo radius, $S(r)$ and $Q(r)$, on the MFs of the semi-analytical haloes. Significant differences remain when compared with the $N$-body haloes, independent on the the few percent bias expected for $S(r)$ and $Q(r)$ \citep{Zemp_2011} as MFs do not depend on any symmetry assumption and parameterisation, further strengthening their interest. However, a possible deviation from the ellipsoidal shape of the $N$-body haloes is not included in the semi-analytical haloes generation method and could be partially responsible for the observed differences. 


\section{Discussion and conclusion} \label{sec: conclu}

A dark-matter halo is usually considered as a spherical distribution of matter following a smooth density profile. This is an oversimplification for several reasons: first, haloes are closer to an ellipsoidal than spherical shapes; second, smooth density profile such as NFW are usually only appropriate for relaxed haloes; finally, as smooth density profiles tend to average out the matter distribution, they erase the information contained in substructures. The morphology of dark-matter haloes is usually described by statistics that are either well-suited for relaxed, ellipsoidal haloes, or empirical expressions that capture specific features of asymmetries but mathematically not well-grounded. In this study, we use the Minkowski Functionals computed using the germ-grain model to capture this complexity.
We generate semi-analytical haloes using parametric density profiles, such as NFW and $\alpha\beta\gamma$-profile as well as non parametric density profiles following the $N$-body haloes particle distribution, and we examine the MFs of these haloes. We use these results as benchmark to interpret the MFs of $N$-body haloes, which share the same average density profile of semi-analytical haloes. Our main conclusions are summarised as follows.

\begin{itemize}
    \item Semi-analytical haloes with parametric profiles allows us to relate the shape of MFs curves to the density profile of haloes, notably to the concentration parameter $c$, logarithmic slopes of inner and outer regions $\gamma$ and $\beta$, sphericity $S$, and elongation $Q$. Higher values of $\beta$ and bu$c$ and lower values of $S$ and $Q$ increase the number of particles in the halo inner region and decrease the number of particles in the halo outer region.
    The inner region has negligible contributions on the MFs of haloes since they are very dense and thus all the particles (the germs) are covered by the balls (the grains) of the neighbouring particles. Consequently, $\gamma$ has only small contributions to the MFs. However, the well-known tensions between observations and simulations of the dark matter distribution in the innermost region of haloes \citep[e.g.][]{De_Blok_2010,Salucci_2019} can be probed by MFs considering specifically central sub-selections of particles; see Appendix \ref{sec: MFs probe different region}.
    Moreover, because of additivity the amplitude of MFs increases with the number of contributing particles, i.e. for lower values of $c$ and $\beta$ and higher $S$ and $Q$. These results are shown in Figures~\ref{fig: MF_sphere_analytic} and~\ref{fig: MF_ellipsoidal_analytic}.
    \item MFs of semi-analytical haloes generated with smooth profiles are significantly different from those of $N$-body haloes, even though the latter are relaxed according to the \citet{Neto_2007} criteria based on virial ratio and centre-of-mass displacement. Although these differences decrease for haloes generated with more realistic profiles, such as ellipsoidal instead of spherical, they do not disappear as caused by deviations from smooth triaxiality, i.e. anisotropies beyond triaxial shape, subhaloes and tidal streams. That is, the MFs can probe more complex morphological features beyond the smooth profiles. This is shown in Figure~\ref{fig: MF real vs SA}, which is the main result of this study.
\end{itemize}

We show that the MFs are a powerful probe to get deep insight into the detailed morphology of haloes, such as subhaloes as shown in Figure~\ref{fig: MF_sub}. Those subhaloes are subject to tidal forces coming from their host halo including also possible spurious numerical effects. As shown in \citet{van_den_Bosch_2018_sim}, artificial mass loss may occur for subhaloes when their half-mass radius is too small compared to the force softening or when they host a too low number of particles.
These criteria are fulfilled for haloes in the most massive bin we have considered, which typically contain $\gtrsim 1000$ particles and have half-mass radius larger than $20 \Delta x$, but not in the intermediate and low mass bins $10^{12.5} < M_\mathrm{vir} h/M_{\odot} <10^{12.7}$ and $10^{13.5} < M_\mathrm{vir} h/M_{\odot} <10^{13.7}$. This is consistent with the deviations between the MFs of the most massive $N$-body haloes and the MFs of their semi-analytic counterpart (see Figure~\ref{fig: MF_SA_vs_DEUS_many}), which are larger than the deviations for haloes in the intermediate and low mass bin (not shown).

A future study aims at investigating the MFs of more realistic semi-analytical haloes containing substructures to weight the contribution of subhaloes, streams or anisotropies beyond triaxial shape. Using subhalo mass functions, the contribution of the clumps with different masses and their impact on the MFs can be addressed. These semi-analytical haloes with subhaloes will be compared with haloes from $N$-body and hydro-dynamical simulations such as The Three Hundred Project \citep{Cui_2018}, Illustris-TNG \citep{Weinberger_2017}, or Horizon-AGN \citep{Dubois_2014}. Such simulations will also enable the study of the connection between the MFs of observable matter (stars, galaxies, or any biased tracer of the underlying matter field) and the MFs of dark matter. A direct application to observational data will require an analogous study in two dimensions possibly including the baryonic component, to finally exploit the accurate projected maps provided by instruments such as eROSITA \citep{Merloni+2020} or Athena \citep{Athena} for X-ray imaging, Vera Rubin Observatory \citep{Ivezic+2019} or Euclid \citep{Laureijs+2011} for gravitational lensing, and Simons Observatory \citep{Hensley+2021} for Sunyaev-Zel'dovich maps. For such studies, the excursion set formalism probing the morphology of the continuous density field \citep{Schmalzing_1997,beisbart2000} is more suitable than the germ-grain model. It will be then interesting to investigate possible relations between the MFs of density maps in two and three dimensions and use MFs to constrain the quality of mass map reconstruction of galaxy clusters.

\section*{Acknowledgements}
The authors are grateful to T.~Buchert and M.~Kerscher for their insightful remarks on Minkowski Functionals, M. Limousin and A.~Nu\~nez-Casti\~neyra for useful discussions, Y.~Rasera and P.~S.~Corasaniti for assistance with DEUS data, J.-C. Lambert for his help on system managing and numerical aspects of the project, and the anonymous referee for the fruitful comments and suggestions.
This research has made use of computing facilities operated by CeSAM data center at LAM, Marseille, France. The Centre de Calcul Intensif d’Aix-Marseille is acknowledged for granting access to its high performance computing resources. CS acknowledges support from the Programme National Cosmology et Galaxies (PNCG) of CNRS/INSU with INP and IN2P3, co-funded by CEA and CNES. KK acknowledges support from the DEEPDIP project (ANR-19-CE31-0023).

\section*{Data Availability}

The data underlying this article will be shared on reasonable request to the corresponding author.



\bibliographystyle{mnras}
\bibliography{mybiblio} 



\appendix


\section{Computational and numerical aspects of MFs}

\subsection{Universality of MFs and characteristic points} \label{subsec: MFs shape}

\begin{figure} \label{fig: MF demo}
 \includegraphics[width=\columnwidth]{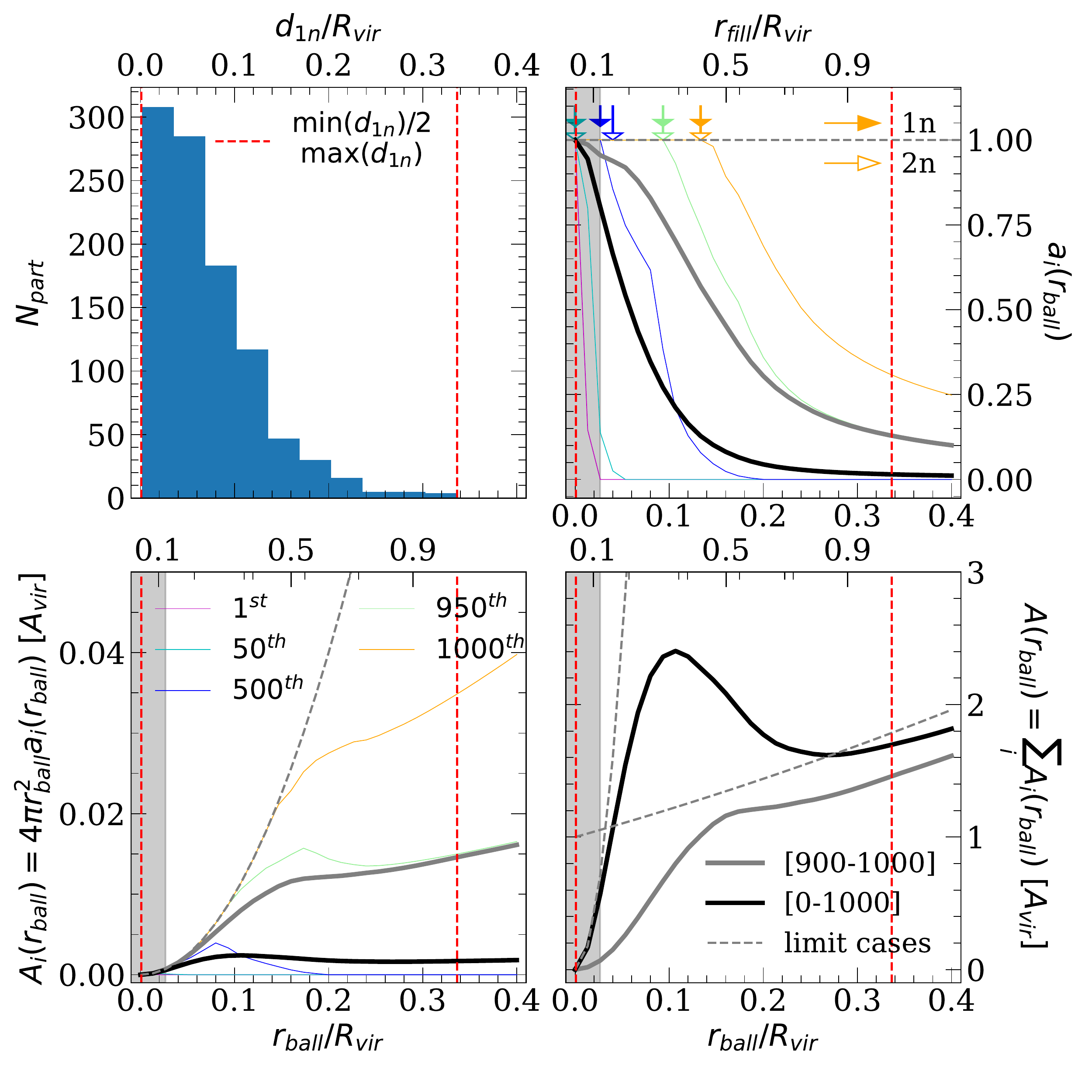}
 \caption{Partial MFs explain the ``universal'' shape: illustration for the area. 
 \textit{Top-left:} Histogram of nearest neighbour distances within one virial radius, $d_\mathrm{1n}$.
 \textit{Bottom-left and top-right:} partial area $A_i$ and dimensionless partial area $a_i$ for the $i$-th particle from the centre (thin coloured lines; $i=1$, 50, 500, 950, 1000). 
 \textit{Bottom-right:} total area.
Black and gray thick curves account for the sum of all particles and the sum of the 100 particles with the largest radii, respectively.
Dashed lines accounts for the limit case of isolated balls and single ball.
Filled (empty) arrows mark the value of $r_\mathrm{ball}$ where the particle starts to intersect with the first (second) neighbours. Resolution limits are indicated by shaded area.
 }
\end{figure}

The MFs of haloes have similar shape, which we refer to as universality, and can be understood in terms of partial MFs \citep{SchmalzingDiaferio2000}. We illustrate this specificity focusing on the area $A$ working on a sample of 1000 particles used in Figure~\ref{fig: MF N_part}; see Figure~\ref{fig: MF demo}.
Particles and corresponding intersecting balls contributing to the area are sorted by their distance from the halo centre. The nearest neighbour distances of all these particles, denoted by $d_\mathrm{1n}$ (histogram in top-left panel), indicates the interesting range for $r_\mathrm{ball}$. The partial area $A_i = 4 \pi r_\mathrm{ball}^2 a_i$ (bottom left panel, thin coloured curves), and its dimensionless counterpart $a_i$
(top-right panel, thin coloured curves), represent respectively the contribution from the particle $i$ to the total area $A$ (bottom-right panel, black curve) and the uncovered normalised solid angle of the ball around the particle $i$. For graphical purposes, we show the contributions for particles $i=1$, 50, 500, 950, and 1000 as sorted by the centre of the halo.

The dimensionless partial MFs $a_i$ is a monotonically decreasing function of the ball radius. It decreases from 1 to 0 as $r_\mathrm{ball}$ increases, except if the $i$-th particle is close to the halo boundary; in this case, it monotonically decreases without reaching $0$ at large ball radius. Correspondingly, for all but the particles at the halo boundary
the partial area $A_i$ increases from 0, reaches a maximum and then asymptotically decreases to $0$ at large ball radius. For the particle close to the halo boundary, $A_i$ continues to increase at a smaller rate at large ball radius, affecting the global trend of $A$. Most of the particles are not at the halo boundary, thus increasing $r_\mathrm{ball}$, the total area $A$ increases from 0, reaches a local maximum and then decreases. At sufficiently large ball radius, the few particles at the halo boundary become the main contribution of the MFs, because the innermost particles are typically covered, carrying vanishing contribution. After the local minimum, $A$ increases again approaching the behaviour of a unique ball, $A=4\pi r_\mathrm{ball}^2$. The area $A$ thus has two extrema. Note also that for large ball radius, the particles at the halo edge (e.g. grey thick curves) contribute to a large fraction of the total MFs (black thick curve in the bottom right panel).

Using similar arguments, one can explain the local minima and maxima of the two other MFs, $H$ and $\chi$.

\subsection{Number of particles} \label{subsec: MF N_part}

\begin{figure} 
 \includegraphics[width=\columnwidth]{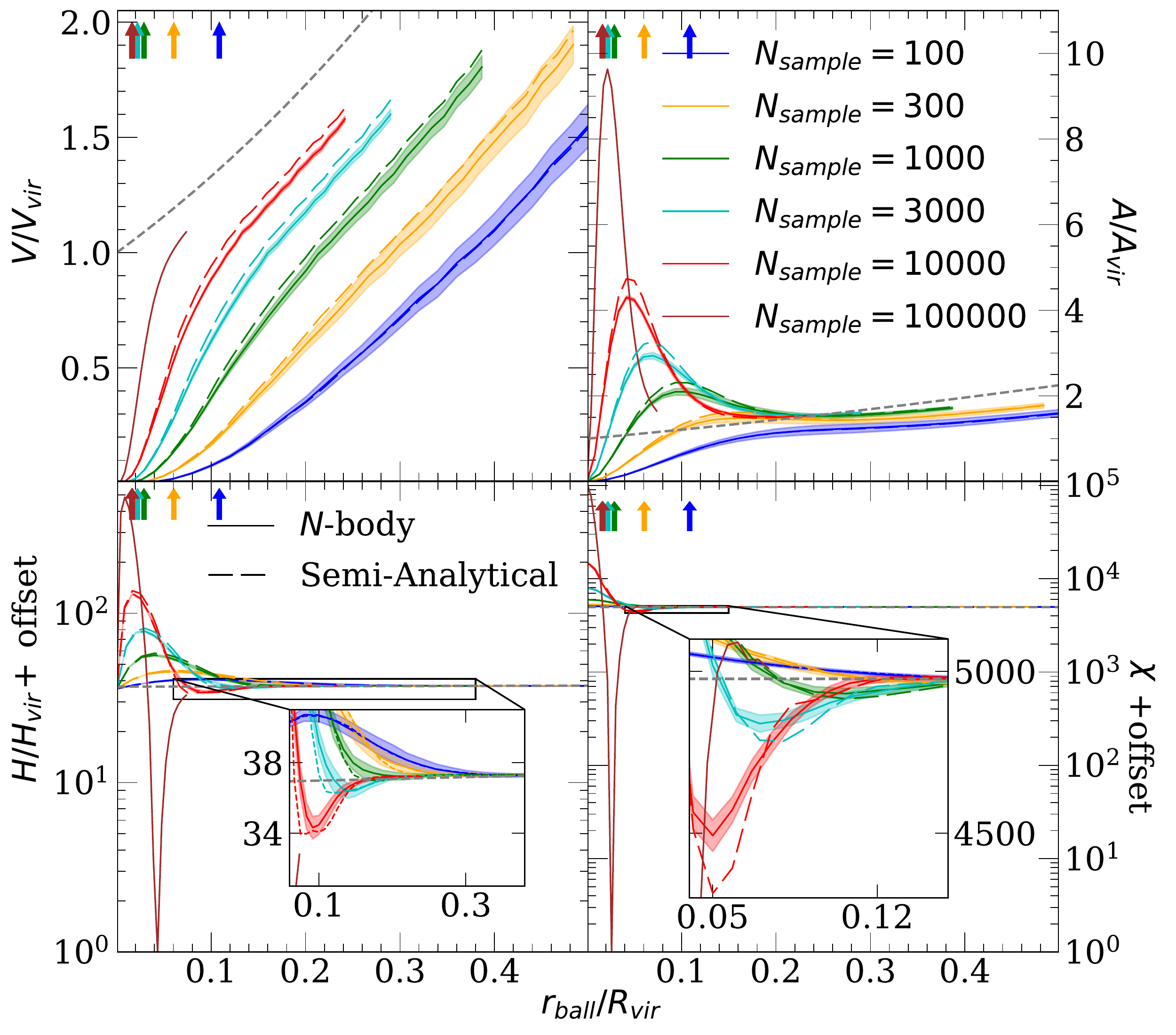}
 \caption{Effect of sampling and internal structures of haloes on MFs, computed for the same haloes shown in Figure~\ref{fig: MF_N_body_SA_N_part}. The overall amplitude and shape of MFs mainly depend on the number of particles $N_\mathrm{sample}$. For all values of $N_\mathrm{sample}$, the internal structures of haloes have a smaller effect, as shown by semi-analytical smooth haloes (long dashed lines) compared to $N$-body haloes (solid lines). The smallness of standard deviations, shown only for the $N$-body halo for clarity, proves the robustness of MFs.
 }
 \label{fig: MF N_part}
\end{figure}

A critical numerical aspect of MFs concerns the dependence on the sample size. Figure~\ref{fig: MF N_part} shows the MFs of the $N$-body and semi-analytical haloes considered in Figure~\ref{fig: MF_N_body_SA_N_part}, sampled with different number of particles $N_\mathrm{sample}$ ranging from $10^2$ to $10^5$. The effect of the internal structures of haloes on the MFs shape is limited. Because of additivity of MFs, their amplitude necessarily increases with $N_\mathrm{sample}$. Moreover, a higher $N_\mathrm{sample}$ shifts the MFs toward lower value of $r_\mathrm{ball}$. This is expected because the nearest neighbour distances $d_{\mathrm{1n},<r}$ and $d_{\mathrm{1n},>r}$ decrease with $N_\mathrm{sample}$. Note also that the local extrema of $A$, $H$ and $\chi$ appear for all the samples. The only exception is the area $A$, which tends to be monotonic for small $N_\mathrm{sample}$. It is important to stress that the level of details of the morphology described by MFs depends on $N_\mathrm{sample}$; however, what matters is the robustness of MFs once a fixed number of particles $N_\mathrm{sample}$ is considered. This is proved by the small standard deviations of MFs (light bands in the figure).

\subsection{Halo filled radius}
\label{section: halo radius probed}

\begin{figure} 
 \includegraphics[width=0.95\columnwidth]{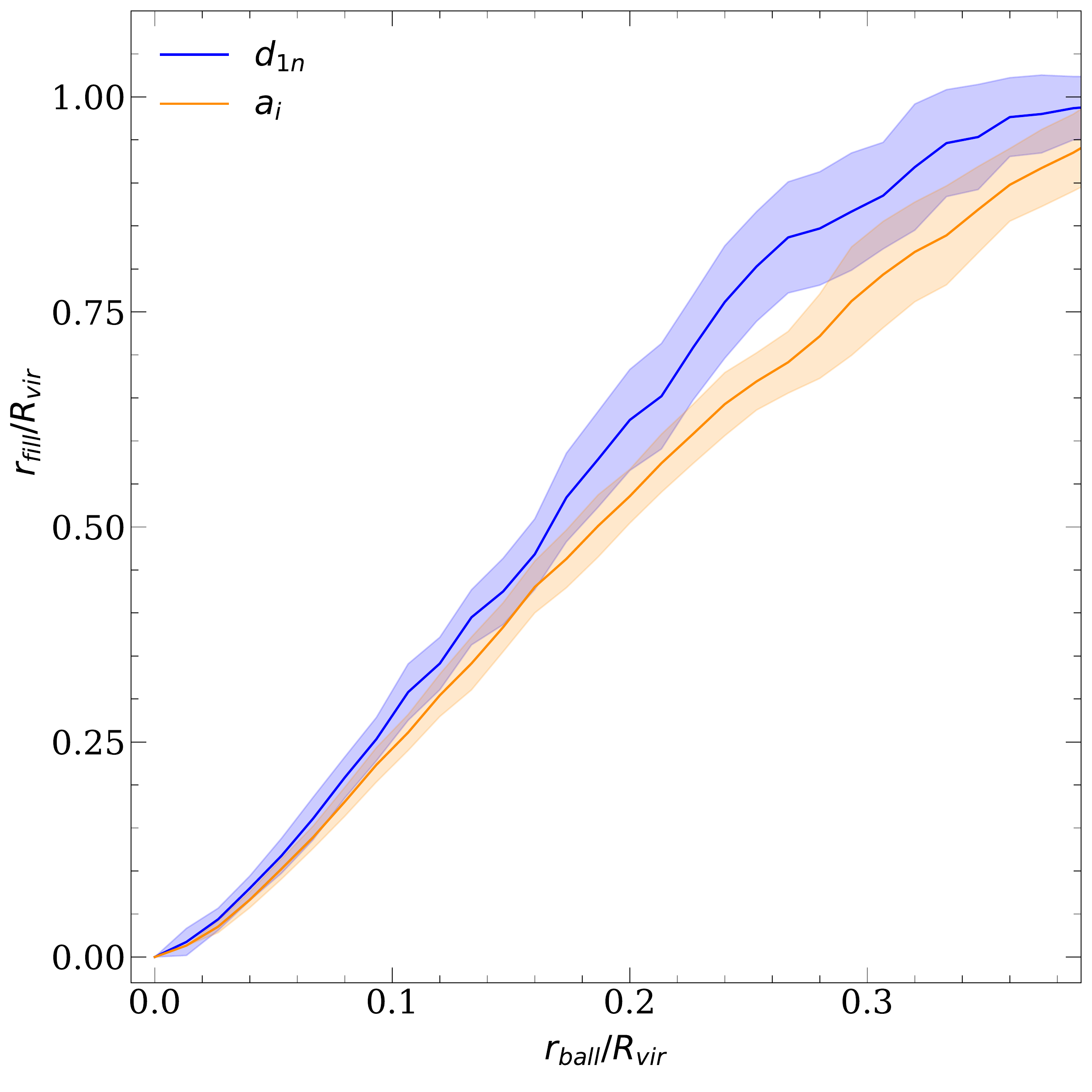}
 \caption{Comparison between $r_\mathrm{fill}$ according the exact Equation~\eqref{eq: r_fill_exact} (orange line) and the approximation Equation~\eqref{eq: max_d1n} (also referred as Equation~\eqref{eq: max_d1n Appendix}, blue line), computed for illustration for the $N$-body halo referring to Figure~\ref{fig: MF N_part} with $N_\mathrm{sample} = 1000$. Error bands are standard deviations computed from 30 samples.
 }
 \label{fig: r_fill_exact}
\end{figure}

The filled radius $r_\mathrm{fill}$ delimits the contributing and non-contributing regions of the halo to the MFs for a given value of $r_\mathrm{ball}$. In this paper, it is approximated using Equation~\eqref{eq: max_d1n}, here recalled for clarity:
\begin{equation} \label{eq: max_d1n Appendix}
r_\mathrm{ball} = \max(d_{\mathrm{1n},<r})\,, \quad\mbox{with}\quad r=r_\mathrm{fill}.
\end{equation}
One can check the accuracy of this equation by computing the exact value of $r_\mathrm{fill}$. For fixed $r_\mathrm{ball}$, the filled radius $r_\mathrm{fill}$ defines the largest sphere of radius $r$ that contains only particles with radius $r_i$ and vanishing dimensionless partial MFs $a_i$, i.e.
\begin{equation} \label{eq: r_fill_exact}
r_\mathrm{fill} = r_\mathrm{ball} + \max \{r: \max\{ a_i: r_i < r\}=0 \}. 
\end{equation}
Figure~\ref{fig: r_fill_exact} shows $r_\mathrm{fill}$ both computed with Equations~(\ref{eq: r_fill_exact}) and (\ref{eq: max_d1n Appendix}) for the halo already considered in Section \ref{subsec: MF N_part}, sampled with 1000 particles.  The two curves of $r_\mathrm{fill}$ show similar trends within the standard deviation, the estimation obtained from the nearest neighbour distances $d_\mathrm{1n}$ (blue curve) being slightly higher than the exact definition based on partial MFs (orange curve). In the current study we therefore adopt the convenient definition based on Equation~\eqref{eq: max_d1n Appendix}.

\subsection{Resolution limit of MFs from sampling} \label{sec: profile_resolution}

\begin{figure} 
\centering
 \includegraphics[width=0.95\columnwidth]{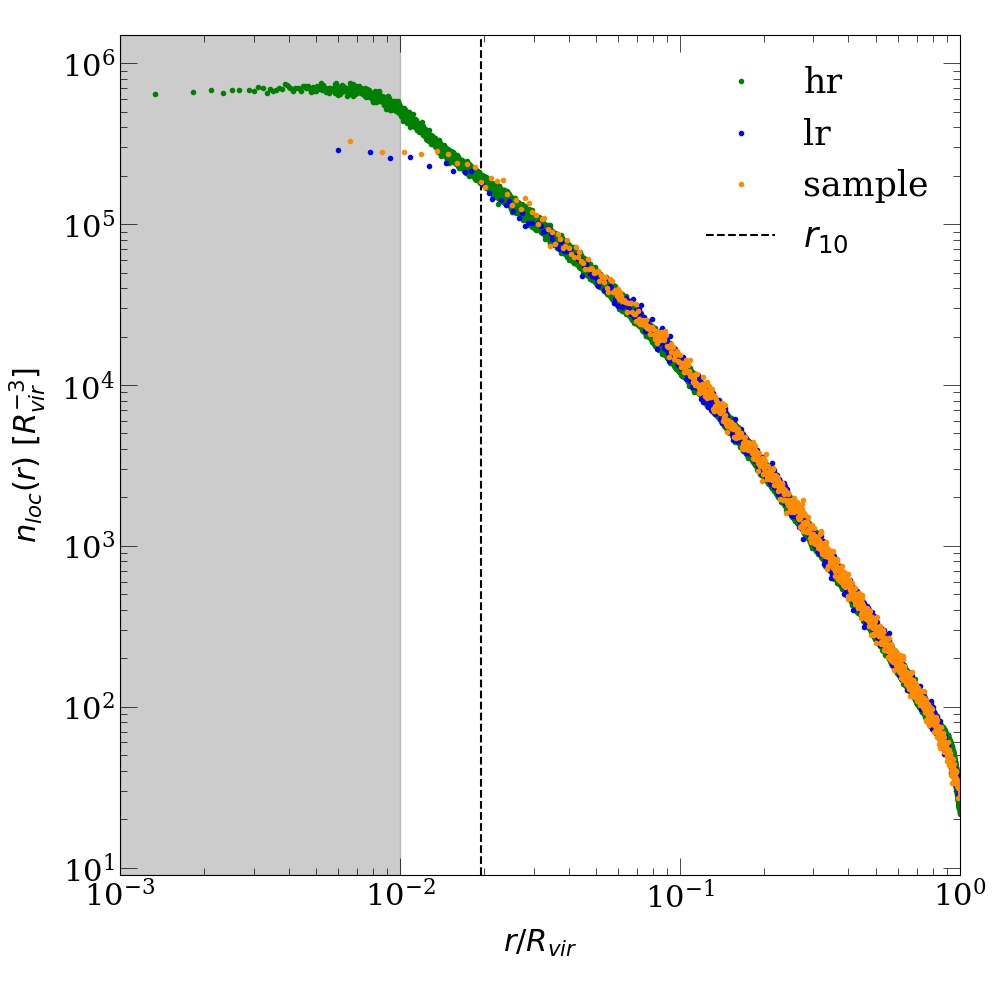}
 \caption{Impact of the sampling on the resolution. Comparison of local density between a first population of $10^5$ particles (green), the sampling of this population with $10^3$ particles (orange), and a second population with $10^3$ particles (blue). All the three haloes share the same $\alpha\beta\gamma$-profile. The unresolved region of the first population is indicated with black shaded area. The radius of the 10-th innermost particle of the sampling $r_{10}$ (vertical dashed line) defines the resolution for the MFs.}
 \label{fig: resolution_profile}
\end{figure}

We investigate here the impact of finite sampling of haloes on resolution, which limits the minimum reliable value of ball radius for the germ-grain model. For this purpose, we shall consider three haloes.
A first halo (or population) of $10^5$ particles is generated with $\alpha\beta\gamma$-profile; its resolution is given by the size of the first shell, which contains about 100 particles. A second halo (or sample) with $10^3$ particles is randomly extracted from the first population; then the size of the first shell containing 100 particles in this sample is larger than for the population. Finally, a third halo is generated with same profile and parameters, but containing only $10^3$ particles. The size of the first shell is the same as the first population, but this shell is not resolved as it contains only $\sim 1$ particle. As shown in Figure~\ref{fig: resolution_profile}, the local density $n_\mathrm{loc}(r)$ of the three processes, i.e. the first population, its sub-sample, and the second population, agrees for $r\gtrsim0.02R_\mathrm{vir}$  radius, and the process with more particles (green points) is more resolved. The sub-sample (orange points) is slightly more resolved than the second population (blue points) though sharing the same number of particles, maybe partially inheriting the resolution of its initial population. The sub-sample is resolved above the radius of 10-th innermost particle, which is therefore used to define a conservative resolution limit $r_\mathrm{res}=r_\mathrm{10}$ for the MFs with Equation~\eqref{eq: res}.

\section{Core-cusp halo Morphology} \label{sec: MFs probe different region}

\begin{figure} 
 \includegraphics[width=\columnwidth]{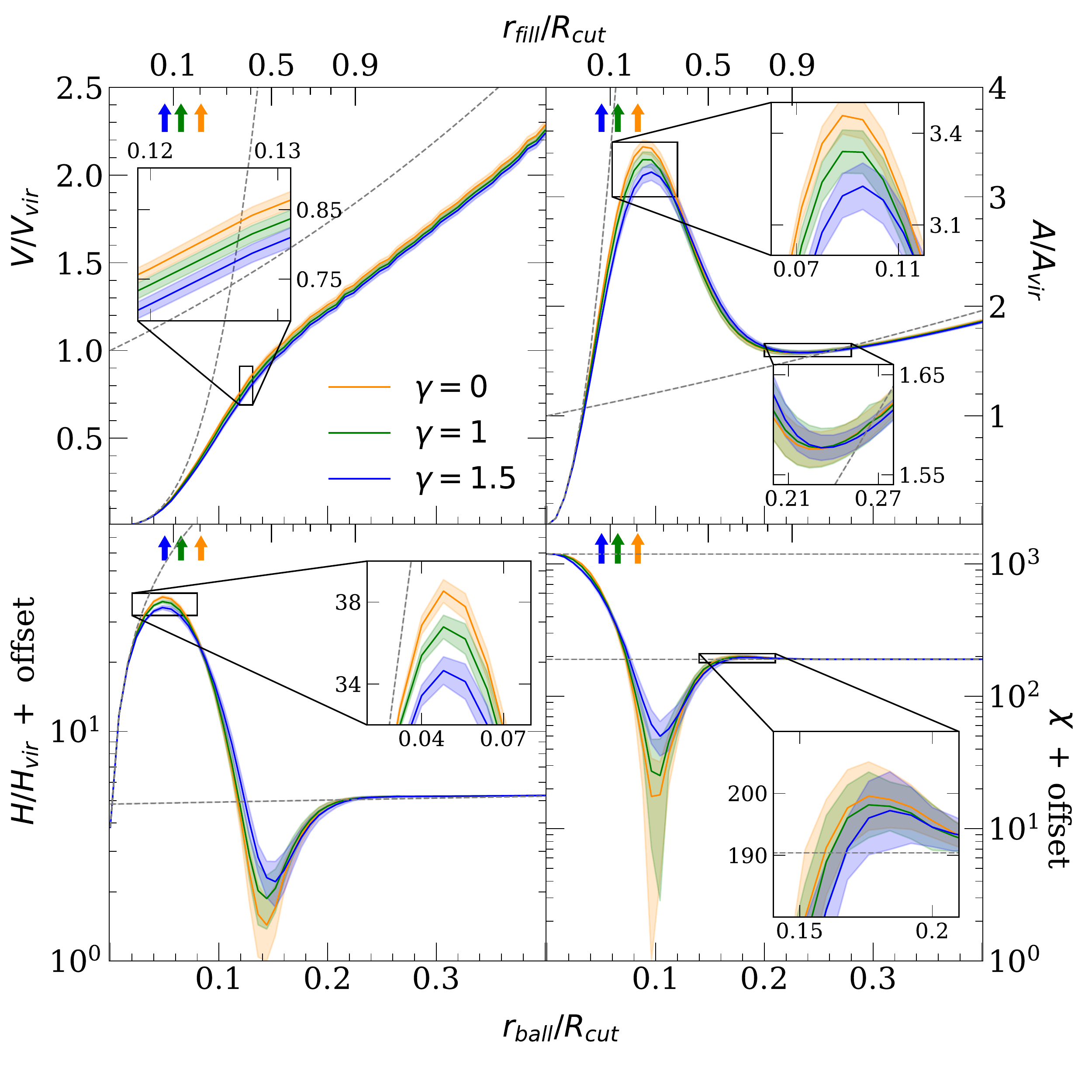}
 \caption{Details of MFs for the halo inner region limited by $R_\mathrm{cut}=0.1R_\mathrm{vir}$ for different values of inner slope $\gamma$. Arrows indicate the resolution computed with Equations~(\ref{eq: res}) and (\ref{eq: res_SA}). To be compared with the fourth column of Figure~\ref{fig: MF_sphere_analytic}.}
 \label{fig: MF gamma inner}
\end{figure}

In this section we provide some additional details about the impact of the inner slope $\gamma$ on the MFs. As discussed in Section~\ref{subsec: SA_spherical}, because of additivity MFs are less sensitive to the inner part of haloes, where multiple covering of balls occurs because of the high density. Still, some morphological information for this region can be obtained by an appropriate selection of points. For this purpose, we generate haloes with the same procedure as in Section~\ref{subsec: halo from profile}, but with a spherical selection $R_\mathrm{cut} = 0.1R_\mathrm{vir}$ before the sampling. The MFs of the resulting processes are shown in Figure \ref{fig: MF gamma inner}.
Higher values of $\gamma$ yield smaller amplitudes of the MFs, contrary to the results shown in the fourth columns of Figure~\ref{fig: MF_sphere_analytic}. In the limit $R_\mathrm{cut}\rightarrow 0$, the processes become equivalent to the scale-free processes, Equation~\eqref{eq: n(r) d}, with $\gamma = \delta$. It also emphasises the potential of the MFs as statistics for the local morphology, as different halo regions can be probed by appropriate selection.

\section{Triaxiality and concentration impact on NFW haloes} \label{sec: MFs_SQ_c}

\begin{figure*} 
 \includegraphics[width=0.9\textwidth]{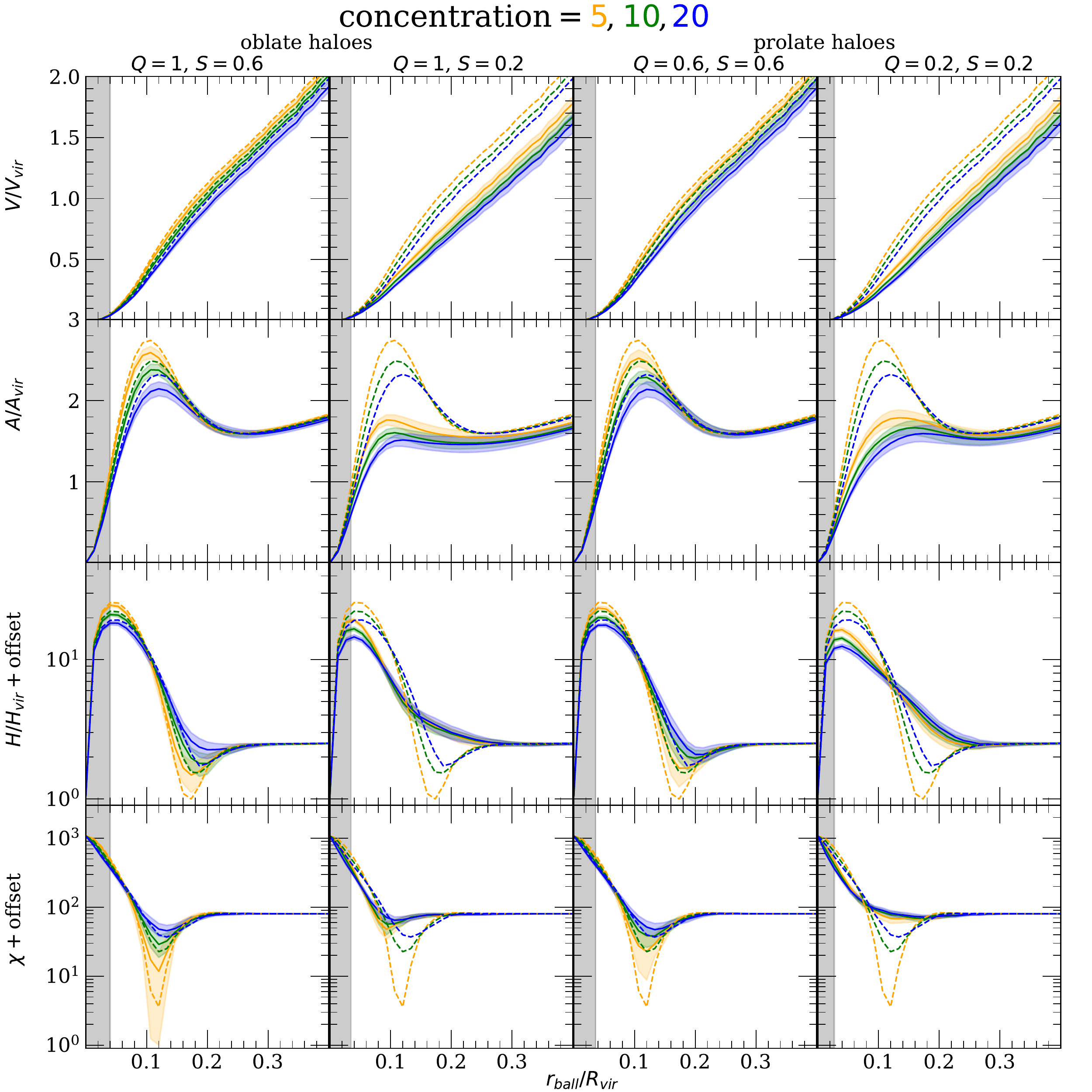}
 \caption{MFs of NFW ellipsoidal haloes generated with the truncated ellipsoidal method with different concentration. Same notations as in Figure~\ref{fig: MF_sphere_analytic}. The mean of the MFs of the NFW spherical haloes of the second column of Figure~\ref{fig: MF_sphere_analytic} are reproduced with dashed lines. The MFs of less spherical and more concentrated haloes have smaller amplitudes.}
 \label{fig: MF_SQ_c}
\end{figure*}

Figure~\ref{fig: MF_SQ_c} shows the impact of both the shape and the concentration of haloes generated with NFW profiles, as a complementary result of Figures~\ref{fig: MF_sphere_analytic}-\ref{fig: MF_ellipsoidal_analytic}. The amplitude of MFs decreases for less spherical and more concentrated haloes.

\section{MFs of semi-analytical haloes with subhaloes} \label{sec: MFs_sub}

\begin{figure} 
 \includegraphics[width=0.9\columnwidth]{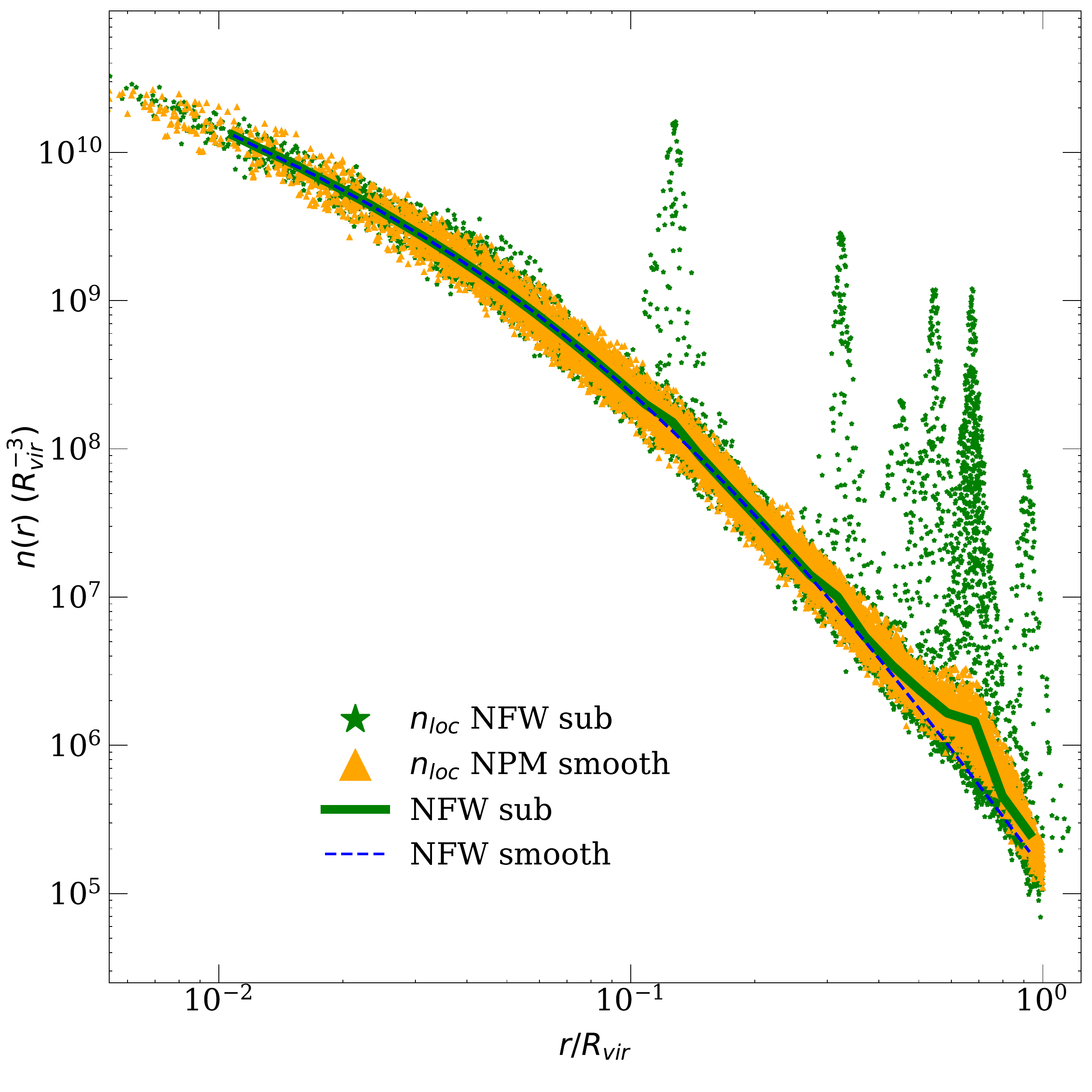}
 \caption{Comparison between three semi-analytical haloes following the notation of Figure~\ref{fig: n_r_n_loc}. Density profile $n(r)$ and local density $n_\mathrm{loc}(\boldsymbol{r})$ of a halo with subhaloes based on the parametric method using a NFW profile (green solid line and symbols). For comparison, $n(r)$ of a halo with the same smooth component (thin blue dashed line) and $n_\mathrm{loc}(\boldsymbol{r})$ of a smooth halo generated with the non-parametric method (``NPM smooth''; orange symbols).}
\label{fig: rho_sub}
\end{figure}

\begin{figure} 
 \includegraphics[width=\columnwidth]{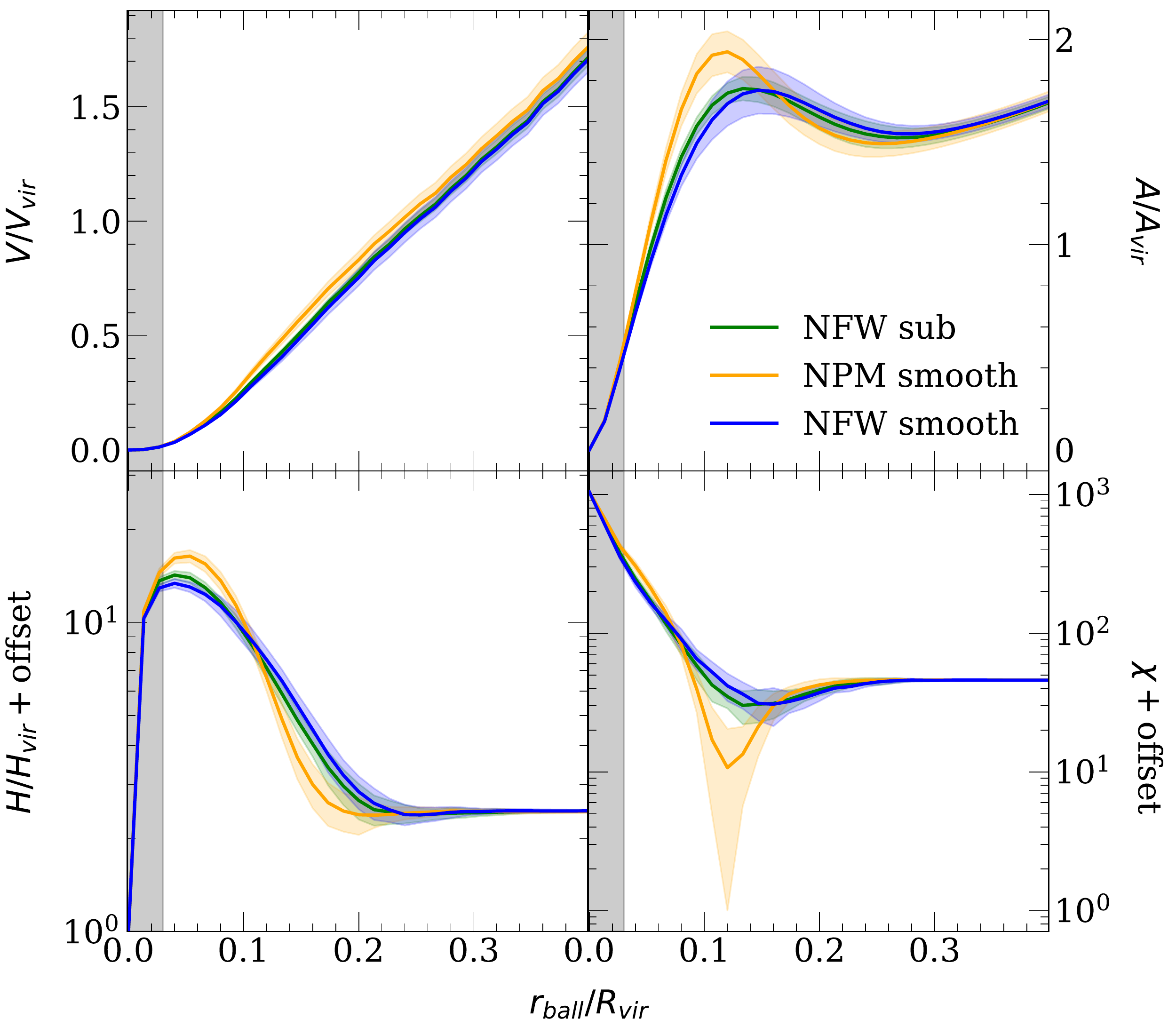}
 \caption{MFs of the three haloes in Figure~\ref{fig: rho_sub}. Although the density profiles of the halo with subhaloes (green) and the smooth non parametric halo (orange) are the same, their MFs are different similarly as in Figure~\ref{fig: MF real vs SA} (orange vs green). In this idealised semi-analytical case, the MFs differences are due to subhaloes.}
\label{fig: MF_sub}
\end{figure}

We have tested the impact of subhaloes, that potentially affect Figures~\ref{fig: MF real vs SA}-\ref{fig: MF_SA_vs_DEUS_many}-\ref{fig: MF_N_body_SA_N_part}, by a toy model based on a simulated semi-analytical halo containing subhaloes. For this purpose, we have generated three spherical semi-analytical haloes: a halo with parametric smooth NFW profile; a second parametric halo with the same smooth component with ten subhaloes of mass $\sim M_\mathrm{main}/100$ added on top, mostly concentrated in the outer region as expected by tidal forces; and a third halo with smooth non-parametric profile sharing the same density profile as the second one. The resulting profiles are shown in Figure~\ref{fig: rho_sub} and the corresponding MFs are shown in Figure~\ref{fig: MF_sub}.

The green and orange lines in Figure~\ref{fig: MF_sub} suggest that the MFs are sensitive to subhaloes. Starting from a smooth NFW profile (Figure~\ref{fig: rho_sub}, blue dashed line), adding subhaloes has less impact on MFs compared to adding the same mass in an overdense shell in the corresponding outer region (compare green and blue lines in Figure~\ref{fig: MF_sub}). Moreover, a smooth halo with less particles in the outer region yields to MFs with lower amplitude (Figure~\ref{fig: MF_sub}, blue vs orange lines), similarly to what is observed in Figure~\ref{fig: MF_sphere_analytic} as an effect of changing $\beta$. Finally, part of the differences observed in Figure~\ref{fig: MF real vs SA}-\ref{fig: MF_SA_vs_DEUS_many}-\ref{fig: MF_N_body_SA_N_part} might still be due to streams and deviation from triaxial shape. A more detailed analysis goes beyond the scope of this paper and will be addressed in a future work.

\bsp	
\label{lastpage}
\end{document}